\documentclass[preprint,12pt,authoryear]{elsarticle}

\usepackage{amssymb}
\usepackage{amsmath}
\usepackage{bm}
\usepackage{url}
\usepackage{hyperref}

\usepackage{changepage}

\usepackage{subcaption}
\usepackage{float}
\usepackage{setspace}

\begin{document}

\begin{frontmatter}

\title{Captions \textemdash\ Assessing the performance of compartmental and renewal models for learning $R_{t}$ using spatially heterogeneous epidemic simulations on real geographies}
\author[label1]{Matthew Ghosh} 
\author[label1]{Yunli Qi}
\author[label1]{Abbie Evans}
\author[label1]{Tom Reed}
\author[label2]{Lara Herriott}
\author[label3]{Ioana Bouros}
\author[label4]{Ben Lambert}
\author[label1]{David J.\ Gavaghan}
\author[label3]{Katherine M.\ Shepherd}
\author[label3]{Richard Creswell\fnref{Richard affiliation}}
\author[label2]{Kit Gallagher}\ead{gallagher@maths.ox.ac.uk}
\affiliation[label1]{organization={Doctoral Training Centre, University of Oxford}, country={UK}}
\affiliation[label2]{organization={Mathematical Institute, University of Oxford}, country={UK}}
\affiliation[label3]{organization={Department of Computer Science, University of Oxford}, country={UK}}
\affiliation[label4]{organization={Department of Statistics, University of Oxford}, country={UK}}

%

\end{frontmatter}

\section*{Figure 1}

{\bf A simulated epidemic across Northern Ireland using Epiabm with an infection radius of 90km.} A: The SEIR compartmental aggregates for our simulation, starting with 100 infected individuals. B: $R_{t}^{\mathrm{case}}$ and the distributions of secondary infections. Each shading represents the number of primary infectors, whose infections started at a specific time point, that go on to have a specific number of secondary infections over the course of their infection. C: The generation time distribution and serial interval distribution, as parameterised within the Epiabm model. D: Offspring distribution for the Epiabm model, and probability mass function for a negative binomial distribution fit to the offspring distribution data between 0 and 9 secondary infections (10 or more secondary infections are suppressed from the plot for visibility). E: Spatial behaviour of the infection wave over time. Each individual frame represents a snapshot in time, where the colour of each pixel represents the total number of infected individuals in that location. The plot was produced using EpiGeoPop \citep{Ellmen_EpiGeoPop}.

\section*{Figure 2}
{\bf SEIR and renewal model inference results from the dataset with an infection radius of 90km.} A: Inferred time series for the SEIR compartmental aggregates with 95\% credible intervals. We sampled 1000 triples ($\beta$, $\kappa$, $\gamma$) from the posterior distribution, simulated the model and calculated the mean, 2.5\textsuperscript{th} percentile and 97.5\textsuperscript{th} percentile to generate the plots. B: The time-varying estimates for the SEIR parameters $\beta$, $\gamma$ and $\kappa$. These are calculated from Eq~\ref{eq:seir-model} using the synthetic compartmental data curves and their derivatives, and compared to true values extracted from Epiabm (see~\nameref{supplementary} Section 6). C: Posterior mean $R_{t}$ curve with 95\% credible intervals for the compartmental model. Using the 1000 triples found in A, we used Eq~\ref{eq:seir-case-Rt} and Eq~\ref{eq:intrins-gen-time} to find $R_{t}$ (using the forward-simulated susceptible curve). We then found the mean and 95\% credible intervals to generate the plots. D: Posterior mean $R_{t}$ curve with 95\% credible intervals for the renewal model.

\section*{Figure 3}
{\bf A simulated epidemic across Northern Ireland using Epiabm with an infection radius of 18km.} A: The SEIR compartmental traces for our simulation, starting with 100 infected individuals --- note the multiple peaks in the infected compartment. B: $R_{t}^{\mathrm{case}}$ and the distributions of secondary infections over time. Each shading represents the number of primary infectors, whose infections started at a specific time point, that go on to have a specific number of secondary infections over the course of their infection. C: Prevalence profiles in different spatial regions of Northern Ireland. D: The geographical locations of the nine spatial regions in panel C. E: Spatial behaviour of the infection wave over time. Each individual frame represents a snapshot in time, where the colour of each pixel represents the total number of infected individuals in that location. Panels D and E were produced using EpiGeoPop \citep{Ellmen_EpiGeoPop}.

\section*{Figure 4}
{\bf SEIR and renewal model inference results from the dataset with an infection radius of 18km.} A: Inferred time series for the SEIR compartments with 95\% credible intervals for the compartmental model. We sampled 1000 triples ($\beta$, $\kappa$, $\gamma$) from the posterior distribution and simulated the model. From here we calculated the mean, 2.5\textsuperscript{th} percentile and 97.5\textsuperscript{th} percentile to generate the plots. B: The time-varying estimates for the SEIR parameters $\beta$, $\gamma$ and $\kappa$. These are calculated from Eq~\ref{eq:seir-model} using the synthetic compartmental data curves and their derivatives (see \nameref{supplementary} Section 6). C: Posterior mean $R_{t}$ curve with 95\% credible interval. Using the 1000 triples found in A, we used the formulae in Eq~\ref{eq:seir-case-Rt} and Eq~\ref{eq:intrins-gen-time} to find $R_{t}$ (using the forward-simulated susceptible curve each time), and plotted the mean and 95\% credible interval. D: Posterior mean $R_{t}$ curve and 95\% credible interval for the renewal model.

\bibliographystyle{elsarticle-harv} 
\bibliography{bibliography}

\begin{thebibliography}{53}
\expandafter\ifx\csname natexlab\endcsname\relax\def\natexlab#1{#1}\fi
\providecommand{\url}[1]{\texttt{#1}}
\providecommand{\href}[2]{#2}
\providecommand{\path}[1]{#1}
\providecommand{\DOIprefix}{}
\providecommand{\ArXivprefix}{arXiv:}
\providecommand{\URLprefix}{URL: }
\providecommand{\Pubmedprefix}{pmid:}
\providecommand{\doi}[1]{\href{http://dx.doi.org/#1}{\path{#1}}}
\providecommand{\Pubmed}[1]{\href{pmid:#1}{\path{#1}}}
\providecommand{\bibinfo}[2]{#2}
\ifx\xfnm\relax \def\xfnm[#1]{\unskip,\space#1}\fi
\bibitem[{Abbott et~al.(2020)Abbott, Hellewell, Thompson, Sherratt, Gibbs, Bosse, Munday, Meakin, Doughty, Chun et~al.}]{abbott2020estimating}
\bibinfo{author}{Abbott, S.}, \bibinfo{author}{Hellewell, J.}, \bibinfo{author}{Thompson, R.N.}, \bibinfo{author}{Sherratt, K.}, \bibinfo{author}{Gibbs, H.P.}, \bibinfo{author}{Bosse, N.I.}, \bibinfo{author}{Munday, J.D.}, \bibinfo{author}{Meakin, S.}, \bibinfo{author}{Doughty, E.L.}, \bibinfo{author}{Chun, J.Y.}, et~al., \bibinfo{year}{2020}.
\newblock \bibinfo{title}{Estimating the time-varying reproduction number of {SARS-CoV-2} using national and subnational case counts}.
\newblock \bibinfo{journal}{Wellcome Open Res.} \bibinfo{volume}{5}, \bibinfo{pages}{112}.
\newblock \DOIprefix\doi{https://doi.org/10.12688/wellcomeopenres.16006.2}.
\bibitem[{Bajaj et~al.(2024)Bajaj, Chen, Creswell, Naidoo, Tsui, Kolade, Nicholson, Lehmann, Hay, Kraemer et~al.}]{bajaj2024covid}
\bibinfo{author}{Bajaj, S.}, \bibinfo{author}{Chen, S.}, \bibinfo{author}{Creswell, R.}, \bibinfo{author}{Naidoo, R.}, \bibinfo{author}{Tsui, J.L.}, \bibinfo{author}{Kolade, O.}, \bibinfo{author}{Nicholson, G.}, \bibinfo{author}{Lehmann, B.}, \bibinfo{author}{Hay, J.A.}, \bibinfo{author}{Kraemer, M.U.}, et~al., \bibinfo{year}{2024}.
\newblock \bibinfo{title}{{COVID}-19 testing and reporting behaviours in {E}ngland across different sociodemographic groups: a population-based study using testing data and data from community prevalence surveillance surveys}.
\newblock \bibinfo{journal}{The Lancet Digit. Health} \bibinfo{volume}{6}, \bibinfo{pages}{E778--E790}.
\newblock \DOIprefix\doi{https://doi.org/10.1016/S2589-7500(24)00169-9}.
\bibitem[{Bansal et~al.(2007)Bansal, Grenfell and Meyers}]{bansal2007individual}
\bibinfo{author}{Bansal, S.}, \bibinfo{author}{Grenfell, B.T.}, \bibinfo{author}{Meyers, L.A.}, \bibinfo{year}{2007}.
\newblock \bibinfo{title}{When individual behaviour matters: homogeneous and network models in epidemiology}.
\newblock \bibinfo{journal}{J. R. Soc. Interface} \bibinfo{volume}{4}, \bibinfo{pages}{879--891}.
\newblock \DOIprefix\doi{https://doi.org/10.1098/rsif.2007.1100}.
\bibitem[{Cairney(2021)}]{Cairney2021}
\bibinfo{author}{Cairney, P.}, \bibinfo{year}{2021}.
\newblock \bibinfo{title}{Evidence-informed {COVID}-19 policy: what problem was the {UK} government trying to solve?}. \bibinfo{publisher}{Edward Elgar Publishing}, \bibinfo{address}{Cheltenham, UK}. chapter~\bibinfo{chapter}{22}.
\newblock pp. \bibinfo{pages}{250--260}.
\newblock \DOIprefix\doi{https://doi.org/10.4337/9781800373594.00034}.
\bibitem[{Cartwright(2017)}]{cartwright2017}
\bibinfo{author}{Cartwright, K.V.}, \bibinfo{year}{2017}.
\newblock \bibinfo{title}{Simpson’s rule cumulative integration with {MS Excel} and irregularly-spaced data}.
\newblock \bibinfo{journal}{Journal of Mathematical Sciences \& Mathematics Education} \bibinfo{volume}{12}, \bibinfo{pages}{1--9}.
\bibitem[{Chadeau-Hyam et~al.(2022)Chadeau-Hyam, Wang, Eales, Haw, Bodinier, Whitaker, Walters, Ainslie, Atchison, Fronterre et~al.}]{chadeau2022sars}
\bibinfo{author}{Chadeau-Hyam, M.}, \bibinfo{author}{Wang, H.}, \bibinfo{author}{Eales, O.}, \bibinfo{author}{Haw, D.}, \bibinfo{author}{Bodinier, B.}, \bibinfo{author}{Whitaker, M.}, \bibinfo{author}{Walters, C.E.}, \bibinfo{author}{Ainslie, K.E.}, \bibinfo{author}{Atchison, C.}, \bibinfo{author}{Fronterre, C.}, et~al., \bibinfo{year}{2022}.
\newblock \bibinfo{title}{{SARS-CoV-2} infection and vaccine effectiveness in {E}ngland ({REACT-1}): a series of cross-sectional random community surveys}.
\newblock \bibinfo{journal}{The Lancet Respir. Med.} \bibinfo{volume}{10}, \bibinfo{pages}{355--366}.
\newblock \DOIprefix\doi{https://doi.org/10.1016/S2213-2600(21)00542-7}.
\bibitem[{Champredon and Dushoff(2015)}]{champredon2015intrinsic}
\bibinfo{author}{Champredon, D.}, \bibinfo{author}{Dushoff, J.}, \bibinfo{year}{2015}.
\newblock \bibinfo{title}{Intrinsic and realized generation intervals in infectious-disease transmission}.
\newblock \bibinfo{journal}{Proceedings of the Royal Society B: Biological Sciences} \bibinfo{volume}{282}, \bibinfo{pages}{20152026}.
\bibitem[{Champredon et~al.(2018)Champredon, Dushoff and Earn}]{Champredon2018}
\bibinfo{author}{Champredon, D.}, \bibinfo{author}{Dushoff, J.}, \bibinfo{author}{Earn, D.J.D.}, \bibinfo{year}{2018}.
\newblock \bibinfo{title}{Equivalence of the {E}rlang-distributed {SEIR} epidemic model and the renewal equation}.
\newblock \bibinfo{journal}{SIAM J. Appl. Math.} \bibinfo{volume}{78}, \bibinfo{pages}{3258--3278}.
\newblock \DOIprefix\doi{https://doi.org/10.1137/18M1186411}.
\bibitem[{Chen et~al.(2020)Chen, Lu, Chang and Liu}]{Chen2020}
\bibinfo{author}{Chen, Y.C.}, \bibinfo{author}{Lu, P.E.}, \bibinfo{author}{Chang, C.S.}, \bibinfo{author}{Liu, T.H.}, \bibinfo{year}{2020}.
\newblock \bibinfo{title}{A time-dependent {SIR} model for {COVID-19} with undetectable infected persons}.
\newblock \bibinfo{journal}{IEEE Trans. Netw. Sci. Eng.} \bibinfo{volume}{7}, \bibinfo{pages}{3279–3294}.
\newblock \DOIprefix\doi{https://doi.org/10.1109/tnse.2020.3024723}.
\bibitem[{Clerx et~al.(2019)Clerx, Robinson, Lambert, Lei, Ghosh, Mirams and Gavaghan}]{Clerx2019Pints}
\bibinfo{author}{Clerx, M.}, \bibinfo{author}{Robinson, M.}, \bibinfo{author}{Lambert, B.}, \bibinfo{author}{Lei, C.L.}, \bibinfo{author}{Ghosh, S.}, \bibinfo{author}{Mirams, G.R.}, \bibinfo{author}{Gavaghan, D.J.}, \bibinfo{year}{2019}.
\newblock \bibinfo{title}{Probabilistic {I}nference on {N}oisy {T}ime {S}eries ({PINTS})}.
\newblock \bibinfo{journal}{J. Open Res. Softw.} \bibinfo{volume}{7}, \bibinfo{pages}{23}.
\newblock \DOIprefix\doi{https://doi.org/10.5334/jors.252}.
\bibitem[{Cori et~al.(2013)Cori, Ferguson, Fraser and Cauchemez}]{cori2013new}
\bibinfo{author}{Cori, A.}, \bibinfo{author}{Ferguson, N.M.}, \bibinfo{author}{Fraser, C.}, \bibinfo{author}{Cauchemez, S.}, \bibinfo{year}{2013}.
\newblock \bibinfo{title}{A new framework and software to estimate time-varying reproduction numbers during epidemics}.
\newblock \bibinfo{journal}{Am. J. Epidemiol.} \bibinfo{volume}{178}, \bibinfo{pages}{1505--1512}.
\newblock \DOIprefix\doi{https://doi.org/10.1093/aje/kwt133}.
\bibitem[{Creswell et~al.(2022)Creswell, Augustin, Bouros, Farm, Miao, Ahern, Robinson, Lemenuel-Diot, Gavaghan, Lambert et~al.}]{creswell2022heterogeneity}
\bibinfo{author}{Creswell, R.}, \bibinfo{author}{Augustin, D.}, \bibinfo{author}{Bouros, I.}, \bibinfo{author}{Farm, H.}, \bibinfo{author}{Miao, S.}, \bibinfo{author}{Ahern, A.}, \bibinfo{author}{Robinson, M.}, \bibinfo{author}{Lemenuel-Diot, A.}, \bibinfo{author}{Gavaghan, D.}, \bibinfo{author}{Lambert, B.}, et~al., \bibinfo{year}{2022}.
\newblock \bibinfo{title}{Heterogeneity in the onwards transmission risk between local and imported cases affects practical estimates of the time-dependent reproduction number}.
\newblock \bibinfo{journal}{Philos. Trans. R. Soc. A} \bibinfo{volume}{380}, \bibinfo{pages}{20210308}.
\newblock \DOIprefix\doi{https://doi.org/10.1098/rsta.2021.0308}.
\bibitem[{Dankwa et~al.(2022)Dankwa, Brouwer and Donnelly}]{Dankwa2022}
\bibinfo{author}{Dankwa, E.A.}, \bibinfo{author}{Brouwer, A.F.}, \bibinfo{author}{Donnelly, C.A.}, \bibinfo{year}{2022}.
\newblock \bibinfo{title}{Structural identifiability of compartmental models for infectious disease transmission is influenced by data type}.
\newblock \bibinfo{journal}{Epidemics} \bibinfo{volume}{41}, \bibinfo{pages}{100643}.
\newblock \DOIprefix\doi{https://doi.org/10.1016/j.epidem.2022.100643}.
\bibitem[{Ellmen et~al.(2024)Ellmen, Bouros and Gallagher}]{Ellmen_EpiGeoPop}
\bibinfo{author}{Ellmen, I.}, \bibinfo{author}{Bouros, I.}, \bibinfo{author}{Gallagher, K.}, \bibinfo{year}{2024}.
\newblock \bibinfo{title}{{EpiGeoPop}}.
\newblock \DOIprefix\doi{https://github.com/SABS-R3-Epidemiology/EpiGeoPop}. \bibinfo{note}{[software]}.
\bibitem[{Farm et~al.(2020)Farm, Miao, Augustin, Gallagher, Ghosh, Fan, Lamirande, Hayman, Bouros, Creswell and Heirene}]{seirmo2020}
\bibinfo{author}{Farm, H.J.}, \bibinfo{author}{Miao, S.}, \bibinfo{author}{Augustin, D.}, \bibinfo{author}{Gallagher, K.}, \bibinfo{author}{Ghosh, M.}, \bibinfo{author}{Fan, N.}, \bibinfo{author}{Lamirande, P.}, \bibinfo{author}{Hayman, E.}, \bibinfo{author}{Bouros, I.}, \bibinfo{author}{Creswell, R.}, \bibinfo{author}{Heirene, L.}, \bibinfo{year}{2020}.
\newblock \bibinfo{title}{{Seirmo}}.
\newblock \URLprefix \url{https://github.com/SABS-R3-Epidemiology/seirmo}. \bibinfo{note}{[software]}.
\bibitem[{Ferguson et~al.(2020)}]{Ferguson2020}
\bibinfo{author}{Ferguson, N.}, et~al., \bibinfo{year}{2020}.
\newblock \bibinfo{title}{Report 9: Impact of non-pharmaceutical interventions ({NPI}s) to reduce {COVID}19 mortality and healthcare demand}.
\newblock \DOIprefix\doi{https://doi.org/10.25561/77482}.
\bibitem[{Gallagher et~al.(2024)Gallagher, Bouros, Fan, Hayman, Heirene, Lamirande, Lemenuel-Diot, Lambert, Gavaghan and Creswell}]{Gallagher2024}
\bibinfo{author}{Gallagher, K.}, \bibinfo{author}{Bouros, I.}, \bibinfo{author}{Fan, N.}, \bibinfo{author}{Hayman, E.}, \bibinfo{author}{Heirene, L.}, \bibinfo{author}{Lamirande, P.}, \bibinfo{author}{Lemenuel-Diot, A.}, \bibinfo{author}{Lambert, B.}, \bibinfo{author}{Gavaghan, D.}, \bibinfo{author}{Creswell, R.}, \bibinfo{year}{2024}.
\newblock \bibinfo{title}{Epidemiological agent-based modelling software ({E}piabm)}.
\newblock \bibinfo{journal}{J. Open Res. Softw.} \bibinfo{volume}{12}, \bibinfo{pages}{3}.
\newblock \DOIprefix\doi{https://doi.org/10.5334/jors.449}.
\bibitem[{Gallagher et~al.(2023)Gallagher, Creswell, Gavaghan and Lambert}]{gallagher2023identification}
\bibinfo{author}{Gallagher, K.}, \bibinfo{author}{Creswell, R.}, \bibinfo{author}{Gavaghan, D.}, \bibinfo{author}{Lambert, B.}, \bibinfo{year}{2023}.
\newblock \bibinfo{title}{Identification and attribution of weekly periodic biases in epidemiological time series data}.
\newblock \bibinfo{journal}{medRxiv} , \bibinfo{pages}{2023--06}\DOIprefix\doi{https://doi.org/10.1101/2023.06.13.23290903}.
\bibitem[{Gelman and Rubin(1992)}]{Gelman1992}
\bibinfo{author}{Gelman, A.}, \bibinfo{author}{Rubin, D.B.}, \bibinfo{year}{1992}.
\newblock \bibinfo{title}{Inference from iterative simulation using multiple sequences}.
\newblock \bibinfo{journal}{Statist. Sci.} \bibinfo{volume}{7}, \bibinfo{pages}{457--472}.
\newblock \DOIprefix\doi{https://doi.org/10.1214/ss/1177011136}.
\bibitem[{Getz et~al.(2019)Getz, Salter and Mgbara}]{getz2019adequacy}
\bibinfo{author}{Getz, W.M.}, \bibinfo{author}{Salter, R.}, \bibinfo{author}{Mgbara, W.}, \bibinfo{year}{2019}.
\newblock \bibinfo{title}{Adequacy of {SEIR} models when epidemics have spatial structure: {E}bola in {S}ierra {L}eone}.
\newblock \bibinfo{journal}{Philos. Trans. R. Soc. B} \bibinfo{volume}{374}, \bibinfo{pages}{20180282}.
\newblock \DOIprefix\doi{https://doi.org/10.1098/rstb.2018.0282}.
\bibitem[{Girardi and Gaetan(2023)}]{girardi2023seir}
\bibinfo{author}{Girardi, P.}, \bibinfo{author}{Gaetan, C.}, \bibinfo{year}{2023}.
\newblock \bibinfo{title}{An {SEIR} model with time-varying coefficients for analyzing the {SARS-CoV-2} epidemic}.
\newblock \bibinfo{journal}{Risk Anal.} \bibinfo{volume}{43}, \bibinfo{pages}{144--155}.
\newblock \DOIprefix\doi{https://doi.org/10.1111/risa.13858}.
\bibitem[{Gostic et~al.(2020)Gostic, McGough, Baskerville, Abbott, Joshi, Tedijanto, Kahn, Niehus, Hay, De~Salazar, Hellewell, Meakin, Munday, Bosse, Sherrat, Thompson, White, Huisman, Scire, Bonhoeffer, Stadler, Wallinga, Funk, Lipsitch and Cobey}]{Gostic2020}
\bibinfo{author}{Gostic, K.M.}, \bibinfo{author}{McGough, L.}, \bibinfo{author}{Baskerville, E.B.}, \bibinfo{author}{Abbott, S.}, \bibinfo{author}{Joshi, K.}, \bibinfo{author}{Tedijanto, C.}, \bibinfo{author}{Kahn, R.}, \bibinfo{author}{Niehus, R.}, \bibinfo{author}{Hay, J.A.}, \bibinfo{author}{De~Salazar, P.M.}, \bibinfo{author}{Hellewell, J.}, \bibinfo{author}{Meakin, S.}, \bibinfo{author}{Munday, J.D.}, \bibinfo{author}{Bosse, N.I.}, \bibinfo{author}{Sherrat, K.}, \bibinfo{author}{Thompson, R.N.}, \bibinfo{author}{White, L.F.}, \bibinfo{author}{Huisman, J.S.}, \bibinfo{author}{Scire, J.}, \bibinfo{author}{Bonhoeffer, S.}, \bibinfo{author}{Stadler, T.}, \bibinfo{author}{Wallinga, J.}, \bibinfo{author}{Funk, S.}, \bibinfo{author}{Lipsitch, M.}, \bibinfo{author}{Cobey, S.}, \bibinfo{year}{2020}.
\newblock \bibinfo{title}{Practical considerations for measuring the effective reproductive number, ${R}_{t}$}.
\newblock \bibinfo{journal}{PLoS Comput. Biol.} \bibinfo{volume}{16}, \bibinfo{pages}{e1008409}.
\newblock \DOIprefix\doi{https://doi.org/10.1371/journal.pcbi.1008409}.
\bibitem[{Green et~al.(2022)Green, Ferguson and Cori}]{Green2022}
\bibinfo{author}{Green, W.D.}, \bibinfo{author}{Ferguson, N.M.}, \bibinfo{author}{Cori, A.}, \bibinfo{year}{2022}.
\newblock \bibinfo{title}{Inferring the reproduction number using the renewal equation in heterogeneous epidemics}.
\newblock \bibinfo{journal}{J. R. Soc. Interface} \bibinfo{volume}{19}, \bibinfo{pages}{20210429}.
\newblock \DOIprefix\doi{https://doi.org/10.1098/rsif.2021.0429}.
\bibitem[{Haario et~al.(2001)Haario, Saksman and Tamminen}]{Haario2001}
\bibinfo{author}{Haario, H.}, \bibinfo{author}{Saksman, E.}, \bibinfo{author}{Tamminen, J.}, \bibinfo{year}{2001}.
\newblock \bibinfo{title}{An adaptive {M}etropolis algorithm}.
\newblock \bibinfo{journal}{Bernoulli} \bibinfo{volume}{7}, \bibinfo{pages}{223--242}.
\newblock \DOIprefix\doi{https://doi.org/10.2307/3318737}.
\bibitem[{Hamilton(1994)}]{Hamilton1994}
\bibinfo{author}{Hamilton, J.D.}, \bibinfo{year}{1994}.
\newblock \bibinfo{title}{Time {S}eries {A}nalysis}. \bibinfo{publisher}{Princeton University Press}, \bibinfo{address}{Princeton}. chapter \bibinfo{chapter}{3.4. Autoregressive Processes}.
\newblock pp. \bibinfo{pages}{53--59}.
\bibitem[{Hansen et~al.(2003)Hansen, Müller and Koumoutsakos}]{Hansen2003}
\bibinfo{author}{Hansen, N.}, \bibinfo{author}{Müller, S.D.}, \bibinfo{author}{Koumoutsakos, P.}, \bibinfo{year}{2003}.
\newblock \bibinfo{title}{Reducing the time complexity of the derandomized evolution strategy with covariance matrix adaptation ({CMA-ES})}.
\newblock \bibinfo{journal}{Evol. Comput.} \bibinfo{volume}{11}, \bibinfo{pages}{1--18}.
\newblock \DOIprefix\doi{https://doi.org/10.1162/106365603321828970}.
\bibitem[{Harris et~al.(2020)Harris, Millman, van~der Walt, Gommers, Virtanen, Cournapeau, Wieser, Taylor, Berg, Smith, Kern, Picus, Hoyer, van Kerkwijk, Brett, Haldane, del R{\'{i}}o, Wiebe, Peterson, G{\'{e}}rard-Marchant, Sheppard, Reddy, Weckesser, Abbasi, Gohlke and Oliphant}]{harris2020array}
\bibinfo{author}{Harris, C.R.}, \bibinfo{author}{Millman, K.J.}, \bibinfo{author}{van~der Walt, S.J.}, \bibinfo{author}{Gommers, R.}, \bibinfo{author}{Virtanen, P.}, \bibinfo{author}{Cournapeau, D.}, \bibinfo{author}{Wieser, E.}, \bibinfo{author}{Taylor, J.}, \bibinfo{author}{Berg, S.}, \bibinfo{author}{Smith, N.J.}, \bibinfo{author}{Kern, R.}, \bibinfo{author}{Picus, M.}, \bibinfo{author}{Hoyer, S.}, \bibinfo{author}{van Kerkwijk, M.H.}, \bibinfo{author}{Brett, M.}, \bibinfo{author}{Haldane, A.}, \bibinfo{author}{del R{\'{i}}o, J.F.}, \bibinfo{author}{Wiebe, M.}, \bibinfo{author}{Peterson, P.}, \bibinfo{author}{G{\'{e}}rard-Marchant, P.}, \bibinfo{author}{Sheppard, K.}, \bibinfo{author}{Reddy, T.}, \bibinfo{author}{Weckesser, W.}, \bibinfo{author}{Abbasi, H.}, \bibinfo{author}{Gohlke, C.}, \bibinfo{author}{Oliphant, T.E.}, \bibinfo{year}{2020}.
\newblock \bibinfo{title}{Array programming with {NumPy}}.
\newblock \bibinfo{journal}{Nature} \bibinfo{volume}{585}, \bibinfo{pages}{357--362}.
\newblock \DOIprefix\doi{https://doi.org/10.1038/s41586-020-2649-2}.
\bibitem[{Heltberg et~al.(2022)Heltberg, Michelsen, Martiny, Christensen, Jensen, Halasa and Petersen}]{heltberg2022spatial}
\bibinfo{author}{Heltberg, M.L.}, \bibinfo{author}{Michelsen, C.}, \bibinfo{author}{Martiny, E.S.}, \bibinfo{author}{Christensen, L.E.}, \bibinfo{author}{Jensen, M.H.}, \bibinfo{author}{Halasa, T.}, \bibinfo{author}{Petersen, T.C.}, \bibinfo{year}{2022}.
\newblock \bibinfo{title}{Spatial heterogeneity affects predictions from early-curve fitting of pandemic outbreaks: a case study using population data from {D}enmark}.
\newblock \bibinfo{journal}{R. Soc. Open Sci.} \bibinfo{volume}{9}, \bibinfo{pages}{220018}.
\newblock \DOIprefix\doi{https://doi.org/10.1098/rsos.220018}.
\bibitem[{Ho et~al.(2023)Ho, Parag, Adam, Lau, Cowling and Tsang}]{ho2023accounting}
\bibinfo{author}{Ho, F.}, \bibinfo{author}{Parag, K.V.}, \bibinfo{author}{Adam, D.C.}, \bibinfo{author}{Lau, E.H.}, \bibinfo{author}{Cowling, B.J.}, \bibinfo{author}{Tsang, T.K.}, \bibinfo{year}{2023}.
\newblock \bibinfo{title}{Accounting for the potential of overdispersion in estimation of the time-varying reproduction number}.
\newblock \bibinfo{journal}{Epidemiology} \bibinfo{volume}{34}, \bibinfo{pages}{201--205}.
\newblock \DOIprefix\doi{https://doi.org/10.1097/EDE.0000000000001563}.
\bibitem[{Hong and Li(2020)}]{Hong2020}
\bibinfo{author}{Hong, H.G.}, \bibinfo{author}{Li, Y.}, \bibinfo{year}{2020}.
\newblock \bibinfo{title}{Estimation of time-varying reproduction numbers underlying epidemiological processes: A new statistical tool for the {COVID-19} pandemic}.
\newblock \bibinfo{journal}{PLoS One} \bibinfo{volume}{15}, \bibinfo{pages}{e0236464}.
\newblock \DOIprefix\doi{https://doi.org/10.1371/journal.pone.0236464}.
\bibitem[{Johnstone et~al.(2016)Johnstone, Chang, Bardenet, de~Boer, Gavaghan, Pathmanathan, Clayton and Mirams}]{Johnstone2016}
\bibinfo{author}{Johnstone, R.H.}, \bibinfo{author}{Chang, E.T.Y.}, \bibinfo{author}{Bardenet, R.}, \bibinfo{author}{de~Boer, T.P.}, \bibinfo{author}{Gavaghan, D.J.}, \bibinfo{author}{Pathmanathan, P.}, \bibinfo{author}{Clayton, R.H.}, \bibinfo{author}{Mirams, G.R.}, \bibinfo{year}{2016}.
\newblock \bibinfo{title}{Uncertainty and variability in models of the cardiac action potential: Can we build trustworthy models?}
\newblock \bibinfo{journal}{J. Mol. Cell. Cardiol.} \bibinfo{volume}{96}, \bibinfo{pages}{49--62}.
\newblock \DOIprefix\doi{http://dx.doi.org/10.1016/j.yjmcc.2015.11.018}.
\bibitem[{Keeling et~al.(2022)Keeling, Dyson, Guyver-Fletcher, Holmes, Semple, Tildesley and Hill}]{Keeling2022}
\bibinfo{author}{Keeling, M.J.}, \bibinfo{author}{Dyson, L.}, \bibinfo{author}{Guyver-Fletcher, G.}, \bibinfo{author}{Holmes, A.}, \bibinfo{author}{Semple, M.G.}, \bibinfo{author}{Tildesley, M.J.}, \bibinfo{author}{Hill, E.M.}, \bibinfo{year}{2022}.
\newblock \bibinfo{title}{Fitting to the {UK} {COVID-19} outbreak, short-term forecasts and estimating the reproductive number}.
\newblock \bibinfo{journal}{Stat. Methods Med. Res.} \bibinfo{volume}{31}, \bibinfo{pages}{1716--1737}.
\newblock \DOIprefix\doi{https://doi.org/10.1177/09622802211070257}.
\bibitem[{Lambert et~al.(2023)Lambert, Lei, Robinson, Clerx, Creswell, Ghosh, Tavener and Gavaghan}]{lambert2023autocorrelated}
\bibinfo{author}{Lambert, B.}, \bibinfo{author}{Lei, C.L.}, \bibinfo{author}{Robinson, M.}, \bibinfo{author}{Clerx, M.}, \bibinfo{author}{Creswell, R.}, \bibinfo{author}{Ghosh, S.}, \bibinfo{author}{Tavener, S.}, \bibinfo{author}{Gavaghan, D.J.}, \bibinfo{year}{2023}.
\newblock \bibinfo{title}{Autocorrelated measurement processes and inference for ordinary differential equation models of biological systems}.
\newblock \bibinfo{journal}{J. R. Soc. Interface} \bibinfo{volume}{20}, \bibinfo{pages}{20220725}.
\newblock \DOIprefix\doi{https://doi.org/10.1098/rsif.2022.0725}.
\bibitem[{Liu et~al.(2021)Liu, Gu and Liu}]{liu2021uncovering}
\bibinfo{author}{Liu, Y.}, \bibinfo{author}{Gu, Z.}, \bibinfo{author}{Liu, J.}, \bibinfo{year}{2021}.
\newblock \bibinfo{title}{Uncovering transmission patterns of {COVID-19} outbreaks: A region-wide comprehensive retrospective study in {H}ong {K}ong}.
\newblock \bibinfo{journal}{EClinicalMedicine} \bibinfo{volume}{36}.
\newblock \DOIprefix\doi{https://doi.org/10.1016/j.eclinm.2021.100929}.
\bibitem[{Lloyd-Smith et~al.(2005)Lloyd-Smith, Schreiber, Kopp and Getz}]{LloydSmith2005}
\bibinfo{author}{Lloyd-Smith, J.O.}, \bibinfo{author}{Schreiber, S.J.}, \bibinfo{author}{Kopp, P.E.}, \bibinfo{author}{Getz, W.M.}, \bibinfo{year}{2005}.
\newblock \bibinfo{title}{Superspreading and the effect of individual variation on disease emergence}.
\newblock \bibinfo{journal}{Nature} \bibinfo{volume}{438}, \bibinfo{pages}{355–359}.
\newblock \URLprefix \url{http://dx.doi.org/10.1038/nature04153}, \DOIprefix\doi{10.1038/nature04153}.
\bibitem[{Mishra et~al.(2020)Mishra, Kwong, Chan and Baral}]{mishra2020understanding}
\bibinfo{author}{Mishra, S.}, \bibinfo{author}{Kwong, J.C.}, \bibinfo{author}{Chan, A.K.}, \bibinfo{author}{Baral, S.D.}, \bibinfo{year}{2020}.
\newblock \bibinfo{title}{Understanding heterogeneity to inform the public health response to {COVID-19} in {C}anada}.
\newblock \bibinfo{journal}{Can. Med. Assoc. J.} \bibinfo{volume}{192}, \bibinfo{pages}{E684--E685}.
\newblock \DOIprefix\doi{https://doi.org/10.1503/cmaj.201112}.
\bibitem[{Nishiura and Chowell(2009)}]{Nishiura2009}
\bibinfo{author}{Nishiura, H.}, \bibinfo{author}{Chowell, G.}, \bibinfo{year}{2009}.
\newblock \bibinfo{title}{The effective reproduction number as a prelude to statistical estimation of time-dependent epidemic trends}. \bibinfo{publisher}{Springer Netherlands}, \bibinfo{address}{Dordrecht}. chapter~\bibinfo{chapter}{5}.
\newblock pp. \bibinfo{pages}{103--121}.
\newblock \DOIprefix\doi{https://doi.org/10.1007/978-90-481-2313-1_5}.
\bibitem[{NISRA(2011)}]{NI_census_2011}
\bibinfo{author}{NISRA}, \bibinfo{year}{2011}.
\newblock \bibinfo{title}{2011 census}.
\newblock \bibinfo{howpublished}{Available at \url{https://www.nisra.gov.uk/statistics/census/2011-census}}.
\newblock \bibinfo{note}{(accessed: 15 May 2024)}.
\bibitem[{Pan et~al.(2020)Pan, Liu, Wang, Guo, Hao, Wang, Huang, He, Yu, Lin, Wei and Wu}]{Pan2020}
\bibinfo{author}{Pan, A.}, \bibinfo{author}{Liu, L.}, \bibinfo{author}{Wang, C.}, \bibinfo{author}{Guo, H.}, \bibinfo{author}{Hao, X.}, \bibinfo{author}{Wang, Q.}, \bibinfo{author}{Huang, J.}, \bibinfo{author}{He, N.}, \bibinfo{author}{Yu, H.}, \bibinfo{author}{Lin, X.}, \bibinfo{author}{Wei, S.}, \bibinfo{author}{Wu, T.}, \bibinfo{year}{2020}.
\newblock \bibinfo{title}{Association of public health interventions with the epidemiology of the {COVID-19} outbreak in {W}uhan, {C}hina}.
\newblock \bibinfo{journal}{JAMA} \bibinfo{volume}{323}, \bibinfo{pages}{1915--1923}.
\newblock \DOIprefix\doi{https://doi.org/10.1001/jama.2020.6130}.
\bibitem[{Parag(2021)}]{parag2021improved}
\bibinfo{author}{Parag, K.V.}, \bibinfo{year}{2021}.
\newblock \bibinfo{title}{Improved estimation of time-varying reproduction numbers at low case incidence and between epidemic waves}.
\newblock \bibinfo{journal}{PLoS Comput. Biol.} \bibinfo{volume}{17}, \bibinfo{pages}{e1009347}.
\newblock \DOIprefix\doi{https://doi.org/10.1371/journal.pcbi.1009347}.
\bibitem[{Parag and Donnelly(2022)}]{Parag2022}
\bibinfo{author}{Parag, K.V.}, \bibinfo{author}{Donnelly, C.A.}, \bibinfo{year}{2022}.
\newblock \bibinfo{title}{Fundamental limits on inferring epidemic resurgence in real time using effective reproduction numbers}.
\newblock \bibinfo{journal}{PLoS Comput. Biol.} \bibinfo{volume}{18}, \bibinfo{pages}{e1010004}.
\newblock \DOIprefix\doi{https://doi.org/10.1371/journal.pcbi.1010004}.
\bibitem[{Parag et~al.(2022a)Parag, Donnelly and Zarebski}]{parag2022quantifying}
\bibinfo{author}{Parag, K.V.}, \bibinfo{author}{Donnelly, C.A.}, \bibinfo{author}{Zarebski, A.E.}, \bibinfo{year}{2022}a.
\newblock \bibinfo{title}{Quantifying the information in noisy epidemic curves}.
\newblock \bibinfo{journal}{Nat. Comput. Sci.} \bibinfo{volume}{2}, \bibinfo{pages}{584--594}.
\newblock \DOIprefix\doi{https://doi.org/10.1038/s43588-022-00313-1}.
\bibitem[{Parag et~al.(2022b)Parag, Thompson and Donnelly}]{parag2022epidemic}
\bibinfo{author}{Parag, K.V.}, \bibinfo{author}{Thompson, R.N.}, \bibinfo{author}{Donnelly, C.A.}, \bibinfo{year}{2022}b.
\newblock \bibinfo{title}{Are epidemic growth rates more informative than reproduction numbers?}
\newblock \bibinfo{journal}{J. R. Stat. Soc., A: Stat. Soc.} \bibinfo{volume}{185}, \bibinfo{pages}{S5--S15}.
\newblock \DOIprefix\doi{https://doi.org/10.1111/rssa.12867}.
\bibitem[{Park et~al.(2020)Park, Champredon and Dushoff}]{Park2020}
\bibinfo{author}{Park, S.W.}, \bibinfo{author}{Champredon, D.}, \bibinfo{author}{Dushoff, J.}, \bibinfo{year}{2020}.
\newblock \bibinfo{title}{Inferring generation-interval distributions from contact-tracing data}.
\newblock \bibinfo{journal}{J. R. Soc. Interface} \bibinfo{volume}{17}, \bibinfo{pages}{20190719}.
\newblock \URLprefix \url{http://dx.doi.org/10.1098/rsif.2019.0719}, \DOIprefix\doi{10.1098/rsif.2019.0719}.
\bibitem[{Rowland et~al.(2021)Rowland, Swannack, Mayo, Parno, Farthing, Dettwiller, George, England, Reif, Cegan, Trump, Linkov, Lafferty and Bridges}]{Rowland2021}
\bibinfo{author}{Rowland, M.A.}, \bibinfo{author}{Swannack, T.M.}, \bibinfo{author}{Mayo, M.L.}, \bibinfo{author}{Parno, M.}, \bibinfo{author}{Farthing, M.}, \bibinfo{author}{Dettwiller, I.}, \bibinfo{author}{George, G.}, \bibinfo{author}{England, W.}, \bibinfo{author}{Reif, M.}, \bibinfo{author}{Cegan, J.}, \bibinfo{author}{Trump, B.}, \bibinfo{author}{Linkov, I.}, \bibinfo{author}{Lafferty, B.}, \bibinfo{author}{Bridges, T.}, \bibinfo{year}{2021}.
\newblock \bibinfo{title}{{COVID-19} infection data encode a dynamic reproduction number in response to policy decisions with secondary wave implications}.
\newblock \bibinfo{journal}{Sci. Rep.} \bibinfo{volume}{11}, \bibinfo{pages}{10875}.
\newblock \DOIprefix\doi{https://doi.org/10.1038/s41598-021-90227-1}.
\bibitem[{Seabold and Perktold(2010)}]{Seabold2010}
\bibinfo{author}{Seabold, S.}, \bibinfo{author}{Perktold, J.}, \bibinfo{year}{2010}.
\newblock \bibinfo{title}{Statsmodels: Econometric and statistical modeling with python}, in: \bibinfo{booktitle}{Proceedings of the 9th Python in Science Conference}, \bibinfo{publisher}{SciPy}. p. \bibinfo{pages}{92–96}.
\newblock \URLprefix \url{http://dx.doi.org/10.25080/Majora-92bf1922-011}, \DOIprefix\doi{10.25080/majora-92bf1922-011}.
\bibitem[{Thompson et~al.(2019)Thompson, Stockwin, {van Gaalen}, Polonsky, Kamvar, Demarsh, Dahlqwist, Li, Miguel, Jombart, Lessler, Cauchemez and Cori}]{Thompson2019}
\bibinfo{author}{Thompson, R.N.}, \bibinfo{author}{Stockwin, J.E.}, \bibinfo{author}{{van Gaalen}, R.D.}, \bibinfo{author}{Polonsky, J.A.}, \bibinfo{author}{Kamvar, Z.N.}, \bibinfo{author}{Demarsh, P.A.}, \bibinfo{author}{Dahlqwist, E.}, \bibinfo{author}{Li, S.}, \bibinfo{author}{Miguel, E.}, \bibinfo{author}{Jombart, T.}, \bibinfo{author}{Lessler, J.}, \bibinfo{author}{Cauchemez, S.}, \bibinfo{author}{Cori, A.}, \bibinfo{year}{2019}.
\newblock \bibinfo{title}{Improved inference of time-varying reproduction numbers during infectious disease outbreaks}.
\newblock \bibinfo{journal}{Epidemics} \bibinfo{volume}{29}, \bibinfo{pages}{100356}.
\newblock \DOIprefix\doi{https://doi.org/10.1016/j.epidem.2019.100356}.
\bibitem[{van~der Vegt et~al.(2022)van~der Vegt, Dai, Bouros, Farm, Creswell, Dimdore-Miles, Cazimoglu, Bajaj, Hopkins, Seiferth, Cooper, Lei, Gavaghan and Lambert}]{van2022learning}
\bibinfo{author}{van~der Vegt, S.A.}, \bibinfo{author}{Dai, L.}, \bibinfo{author}{Bouros, I.}, \bibinfo{author}{Farm, H.J.}, \bibinfo{author}{Creswell, R.}, \bibinfo{author}{Dimdore-Miles, O.}, \bibinfo{author}{Cazimoglu, I.}, \bibinfo{author}{Bajaj, S.}, \bibinfo{author}{Hopkins, L.}, \bibinfo{author}{Seiferth, D.}, \bibinfo{author}{Cooper, F.}, \bibinfo{author}{Lei, C.L.}, \bibinfo{author}{Gavaghan, D.}, \bibinfo{author}{Lambert, B.}, \bibinfo{year}{2022}.
\newblock \bibinfo{title}{Learning transmission dynamics modelling of covid-19 using comomodels}.
\newblock \bibinfo{journal}{Math. Biosci.} \bibinfo{volume}{349}, \bibinfo{pages}{108824}.
\newblock \DOIprefix\doi{https://doi.org/10.1016/j.mbs.2022.108824}.
\bibitem[{Viboud et~al.(2006)Viboud, Bj{\o}rnstad, Smith, Simonsen, Miller and Grenfell}]{viboud2006synchrony}
\bibinfo{author}{Viboud, C.}, \bibinfo{author}{Bj{\o}rnstad, O.N.}, \bibinfo{author}{Smith, D.L.}, \bibinfo{author}{Simonsen, L.}, \bibinfo{author}{Miller, M.A.}, \bibinfo{author}{Grenfell, B.T.}, \bibinfo{year}{2006}.
\newblock \bibinfo{title}{Synchrony, waves, and spatial hierarchies in the spread of influenza}.
\newblock \bibinfo{journal}{Science} \bibinfo{volume}{312}, \bibinfo{pages}{447--451}.
\newblock \DOIprefix\doi{https://doi.org/10.1126/science.1125237}.
\bibitem[{Wallinga and Lipsitch(2007)}]{Wallinga2007}
\bibinfo{author}{Wallinga, J.}, \bibinfo{author}{Lipsitch, M.}, \bibinfo{year}{2007}.
\newblock \bibinfo{title}{How generation intervals shape the relationship between growth rates and reproductive numbers}.
\newblock \bibinfo{journal}{Proc. R. Soc. B Biol. Sci.} \bibinfo{volume}{274}, \bibinfo{pages}{599--604}.
\newblock \DOIprefix\doi{https://doi.org/10.1098/rspb.2006.3754}.
\bibitem[{You et~al.(2020)You, Deng, Hu, Sun, Lin, Zhou, Pang, Zhang, Chen and Zhou}]{You2020}
\bibinfo{author}{You, C.}, \bibinfo{author}{Deng, Y.}, \bibinfo{author}{Hu, W.}, \bibinfo{author}{Sun, J.}, \bibinfo{author}{Lin, Q.}, \bibinfo{author}{Zhou, F.}, \bibinfo{author}{Pang, C.H.}, \bibinfo{author}{Zhang, Y.}, \bibinfo{author}{Chen, Z.}, \bibinfo{author}{Zhou, X.H.}, \bibinfo{year}{2020}.
\newblock \bibinfo{title}{Estimation of the time-varying reproduction number of {COVID}-19 outbreak in {C}hina}.
\newblock \bibinfo{journal}{Int. J. Hyg. Environ. Health} \bibinfo{volume}{228}, \bibinfo{pages}{113555}.
\newblock \DOIprefix\doi{https://doi.org/10.1016/j.ijheh.2020.113555}.
\bibitem[{Zelenkov and Reshettsov(2023)}]{zelenkov2023analysis}
\bibinfo{author}{Zelenkov, Y.}, \bibinfo{author}{Reshettsov, I.}, \bibinfo{year}{2023}.
\newblock \bibinfo{title}{Analysis of the {COVID-19} pandemic using a compartmental model with time-varying parameters fitted by a genetic algorithm}.
\newblock \bibinfo{journal}{Expert Syst. Appl.} \bibinfo{volume}{224}, \bibinfo{pages}{120034}.
\newblock \DOIprefix\doi{https://doi.org/10.1016/j.eswa.2023.120034}.
\bibitem[{Zhou et~al.(2022)Zhou, Kolaczyk, Thompson and White}]{zhou2022estimation}
\bibinfo{author}{Zhou, Z.}, \bibinfo{author}{Kolaczyk, E.D.}, \bibinfo{author}{Thompson, R.N.}, \bibinfo{author}{White, L.F.}, \bibinfo{year}{2022}.
\newblock \bibinfo{title}{Estimation of heterogeneous instantaneous reproduction numbers with application to characterize {SARS-CoV-2} transmission in {M}assachusetts counties}.
\newblock \bibinfo{journal}{PLOS Comput. Biol.} \bibinfo{volume}{18}, \bibinfo{pages}{e1010434}.
\newblock \DOIprefix\doi{https://doi.org/10.1371/journal.pcbi.1010434}.

\end{thebibliography}

\end{document}


\begin{frontmatter}



\title{Assessing the performance of compartmental and renewal models for learning $R_{t}$ using spatially heterogeneous epidemic simulations on real geographies}

\author[label1]{Matthew Ghosh} 
\author[label1]{Yunli Qi}
\author[label1]{Abbie Evans}
\author[label1]{Tom Reed}
\author[label2]{Lara Herriott}
\author[label3]{Ioana Bouros}
\author[label4]{Ben Lambert}
\author[label1]{David J.\ Gavaghan}
\author[label3]{Katherine M.\ Shepherd}
\author[label3]{Richard Creswell\fnref{Richard affiliation}}
\author[label2]{Kit Gallagher}
\affiliation[label1]{organization={Doctoral Training Centre, University of Oxford}, country={UK}}
\affiliation[label2]{organization={Mathematical Institute, University of Oxford}, country={UK}}
\affiliation[label3]{organization={Department of Computer Science, University of Oxford}, country={UK}}
\affiliation[label4]{organization={Department of Statistics, University of Oxford}, country={UK}}


\begin{abstract}The time-varying reproduction number ($R_t$) gives an indication of the trajectory of an infectious disease outbreak. Commonly used frameworks for inferring  $R_t$ from epidemiological time series include those based on compartmental models (such as the SEIR model) and renewal equation models. These inference methods are usually validated using synthetic data generated from a simple model, often from the same class of model as the inference framework. However, in a real outbreak the transmission processes, and thus the infection data collected, are much more complex. The performance of common $R_t$ inference methods on data with similar complexity to real world scenarios has been subject to less comprehensive validation. We therefore propose evaluating these inference methods on outbreak data generated from a sophisticated, geographically accurate agent-based model. We illustrate this proposed method by generating synthetic data for two outbreaks in Northern Ireland --- one with minimal spatial heterogeneity, and one with additional heterogeneity.
We find that the simple SEIR model struggles with the greater heterogeneity, while the renewal equation model demonstrates greater robustness to spatial heterogeneity, though is sensitive to the accuracy of the generation time distribution used in inference. 
Our approach represents a principled way to benchmark epidemiological inference tools and 
is built upon an open-source software platform for reproducible epidemic simulation and inference.


\end{abstract}

\begin{graphicalabstract}
\centering
\includegraphics[width=\linewidth]{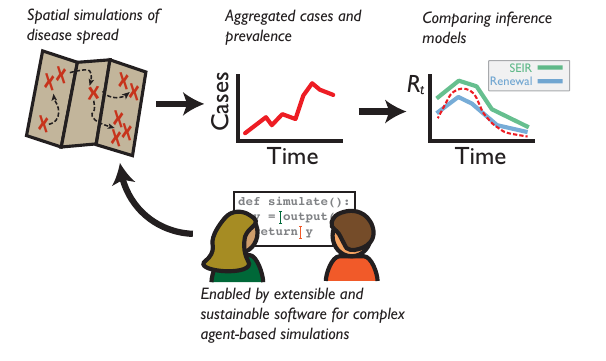}
\end{graphicalabstract}

\begin{highlights}
\item Agent-based models (ABMs) can generate data for complex outbreaks
\item We can calculate the `ground-truth' reproduction number ($R_t$) for these ABM data
\item The ABM data and `ground-truth' $R_t$ can be used to evaluate methods for inferring $R_t$
\item SEIR and renewal models can infer $R_t$ for ABM data with little spatial variation
\item Simple SEIR models may struggle with increased spatial variation in the data
\end{highlights}

\begin{keyword}
Infectious disease epidemiology \sep Agent-based modelling \sep Bayesian inference \sep Reproduction number \sep Model validation \sep Spatial heterogeneity \sep Model complexity 

\end{keyword}

\end{frontmatter}


\section{Introduction}

During a disease outbreak, the ability to predict future cases, to assess the impact of interventions, or to characterize the transmissibility of the disease is typically contingent on analysing different epidemiological time series, such as cases, infection prevalence, or deaths, sometimes in combination with other data about the pathogen and the population it is infecting. But both the time series, and the models typically fit to them, do not include full information about the individual characteristics of each infected person and transmission event. For example, case numbers are often aggregated across geographical regions. Even when more detailed data are available, such as the specific location or age of each individual, the types of models that can accurately incorporate this information tend to be complex, and therefore computationally expensive to fit. Some of the most complex, and arguably realistic, models of disease outbreaks include agent-based models (ABMs), and these may have thousands of tunable parameters. The large number of parameters, combined with runtimes of minutes or hours for a single simulation, typically renders ABMs unsuitable for inferring epidemiological parameters. Instead, simpler models are often used to analyse epidemiological time series, although these rarely capture the heterogeneities of real disease outbreaks, in which transmission is known to be affected by spatial location, demographics, age, travel history and social networks~\citep{LloydSmith2005, viboud2006synchrony, mishra2020understanding, liu2021uncovering, zhou2022estimation, bajaj2024covid}.

These simpler models can be used to infer the instantaneous time-varying reproduction number, $R_t$, which gives the average number of secondary cases each person infected with the disease will go on to cause, assuming transmission remains constant \citep{Nishiura2009, Gostic2020}. $R_t$ is useful for interpreting the trajectory of an outbreak: if $R_t < 1$ an epidemic will eventually cease, but if $R_t > 1$, then it will grow. $R_t$ was therefore inferred, and used extensively, as part of policymaking during the COVID-19 pandemic \citep{Pan2020, Cairney2021, Rowland2021}. 

Two widespread families of methods for learning $R_t$ include those based on compartmental differential equations, such as the SEIR paradigm (e.g. \citet{van2022learning}), and those based on discrete renewal equations \citep{cori2013new, Thompson2019}. While these approaches may be applied directly to real world data \citep{You2020, Green2022}, their validation requires synthetic data so that inferred $R_{t}$ values may be compared against a known ground truth. 

However, the synthetic datasets and underlying models used in prior work typically lack the complexity of real-world epidemic data. For example, \citet{Gostic2020} generated data from deterministic and stochastic ODE models and compared the accuracy of three different $R_{t}$ estimation methods. \citet{Parag2022} used Bayesian inference to infer the value of $R_{t}$ given the generation time and incidence data simulated by a renewal model of an Ebola epidemic; this model considered arbitrary distinct groups to emulate spatial or demographic heterogeneities, though these were not based on geographical data or behaviours of different age groups. 

\citet{bansal2007individual} considered the suitability of the compartmental differential equation approach to inferring $R_{t}$ from data generated by heterogenous processes, proposing network-based modelling, or alternatively modifications of the conventional compartmental equations, to relax the assumption of homogenous mixing. Also starting with a standard SEIR model, \citet{getz2019adequacy} accounted for some proportion of the population being unsusceptible by inferring a parameter for the population-at-risk. The results from this modified SEIR model showed that parameter values inferred from country-wide spatially aggregated data on the 2014--2015 outbreak of Ebola in Sierra Leone deviated from those inferred based on finer, district-level fits of the same model.

Building on the approach taken by \citet{heltberg2022spatial}, which evaluated the ability of a conventional non-spatial SEIR model to predict cases simulated from a spatially structured model, in this study we generate synthetic data from a more sophisticated agent-based model and compare a wider range of different methods for inferring $R_t$. We utilise an agent-based software called Epiabm developed by \citet{Gallagher2024}, based on the CovidSim model from Imperial College London \citep{Ferguson2020}. This offers a complex computational model with age and spatial heterogeneity, and simulates disease transmission dynamics on a national scale. Specifically, we simulate the spread of a COVID-like disease in Northern Ireland, parameterised using region-specific geospatial and demographic (census) data.

We then contrast two standard classes of approaches to $R_{t}$ inference: an SEIR differential equation system and a renewal model. We find that with highly spatially heterogeneous data, the renewal model is better able to capture the $R_{t}$ curve. Such benchmarking could usefully improve the inference of epidemiological parameters such as $R_{t}$ following a future outbreak of an infectious disease, in turn helping decision making around policy by ensuring more accurate information on the current stage of an outbreak is available.

This work is accompanied by open-source notebooks and scripts to reproduce all steps of the analysis.

\section{Methods}
\subsection{Agent-based model --- synthetic data} \label{sec:agent-based-model}

Agent-based models (ABMs) explicitly model people as individual agents with distinct characteristics and behaviour. Epidemiological ABMs simulate both the progression of an infection within an individual, as well as the transmission of the disease between individuals. The ABM software Epiabm \citep{Gallagher2024} is based on an extension of the compartmental SEIRD (Susceptible/Exposed/Infectious/Recovered/Deceased) model, and models infections separately within households, in schools/workplaces, and via wider social interactions. Further details of the ABM are given by \citet{Gallagher2024}.

We generated synthetic data for two outbreaks in Northern Ireland using Epiabm and the accompanying software package EpiGeoPop~\citep{Ellmen_EpiGeoPop}. First, we used EpiGeoPop to generate the geospatial data for the country. This included the number of households, places (such as schools and care homes) and the individuals assigned to each geographical region. The distribution of the number of occupants per household was informed by data from the 2011 Census~\citep{NI_census_2011}. We then used Epiabm to simulate the two outbreaks, only varying the infection radius parameter between the two simulations. This parameter is defined to be the maximal allowed distance between an infector and their potential infectee, measured between the centroids of the regions containing the infector and potential infectee, and is specified in degrees (in the simulation location, 1 degree is equivalent to $90$ kilometres). Information on additional important parameters used in the simulations are provided in~\nameref{supplementary} Section 1.

We extended the Epiabm package so that we were able to collect extra information from the two ABM simulations. Tracking the case reproduction number over time gave us a `ground truth' against which to compare $R_t$ values inferred using the SEIR and renewal model inference frameworks. To compute daily $R_{t}$ values, we assigned the number of secondary infections an individual went on to make to the day in which they first entered the exposed compartment. This count was averaged across all individuals who were exposed on a given day to determine the $R_{t}$ value for that day. We linearly interpolated the value of $R_{t}$ for the days in which no one entered the exposed compartment (and hence no secondary infections were recorded).

We also extracted serial intervals, generation times and offspring distributions (i.e.\ distributions for the number of secondary infections that each infected individual causes) from the two ABM simulations. We assumed that, for symptomatic individuals, symptom onset occurred at the time of entry into the infectious compartment. We therefore extracted distributions for the  number of days between a primary case and a secondary case entering one of the infected compartments in the ABM. We note, however, that these are not formal serial intervals, as we included asymptomatic individuals in this calculation. The generation times were extracted similarly, and were defined as the number of days between primary and secondary cases entering the exposed compartment. Finally, prevalence and incidence data were calculated from time series data containing individual compartmental counts from the ABM.

We fit the extracted offspring distributions to negative binomial distributions using a maximum likelihood approach, as implemented in the package statsmodels~\citep{Seabold2010}. The negative binomial distribution provides a good fit for offspring distributions extracted from real data \citep{LloydSmith2005}, so this provided insight into the realism of the Epiabm simulations. It also allowed us to compare the properties of the two ABM simulations.

\subsection{SEIR model and inference framework}

\subsubsection{Formalism of the ODE system}

The compartmental SEIR model for a total population of size $N$ is as follows:

\begin{subequations} \label{eq:seir-model}
\begin{align}
    \frac{dS(t)}{dt} &= -\frac{\beta}{N} S(t)I(t),\\
    \frac{dE(t)}{dt} &= \frac{\beta}{N} S(t)I(t) - \kappa E(t),\\
    \frac{dI(t)}{dt} &= \kappa E(t) - \gamma I(t),\\
    \frac{dR(t)}{dt} &= \gamma I(t),
\end{align}
\end{subequations}

\noindent for (S)usceptible, (E)xposed, (I)nfected and (R)ecovered compartments, where $\beta$ is the transmission parameter, $\kappa$ is the incubation rate and $\gamma$ is the recovery rate. We used the Seirmo package~\citep{seirmo2020} from the SABS-R3-Epidemiology community to simulate and perform inference using this model. The model is simulated using an explicit Runge-Kutta method of order 5(4), with an adaptive time step. Whilst all compartment populations are tracked within the ODE model, only the population in the infected compartment is used in inference.

\subsubsection{Computation of the instantaneous and case reproduction numbers from the ODE system} \label{sec:reproduction-number}
The instantaneous reproduction number, $R_{t}^{\mathrm{inst}}$, can be easily extracted from the SEIR ODE system and represents the transmission of disease at a specific time. However, this is not equivalent to the reproduction number recorded in Epiabm, which is instead the case reproduction number, $R_{t}^{\mathrm{case}}$. This number tracks transmission based on groups of individuals with the same time of infection, $t$~\citep{cori2013new}.

To estimate the instantaneous reproduction number from the SEIR system, we used the following relationship from \citet{Gostic2020}:

\begin{equation} \label{eq:seir-instantaneous-Rt}
	R_{t}^{\mathrm{inst}} = \frac{\beta}{\gamma}\frac{S(t)}{N}.
\end{equation}

The time dependence of $R_{t}^{\mathrm{inst}}$ is solely determined by the number of susceptible individuals present in the population, i.e.\ $R_{t}^{\mathrm{inst}} \propto S(t)$ when using an SEIR model. To convert to the case reproduction number as extracted from Epiabm, we adopted the approach by \citet{Gostic2020}, according to the methodology first presented by \citet{Wallinga2007}:

\begin{equation} \label{eq:seir-case-Rt}
	R_{t}^{\mathrm{case}} = \int_{t}^{T}R_{u}^{\mathrm{inst}}f(u - t)du,
\end{equation}

\noindent for the case reproduction number, $R_{t}^{\mathrm{case}}$, and the duration of the simulation, $T$. We compute this integral numerically, using the Simpson method \citep{cartwright2017}, though it may also be possible to reduce this to a summation when $R_{u}^{\mathrm{inst}}$ is a discrete time series.

Additionally, the intrinsic distribution of generation times, $f(t)$, is defined as the probability density that the generation time between a primary and secondary infection is equal to $t$ days, assuming a fully susceptible population (see e.g.\ \citep{Park2020}). The intrinsic distribution of generation times for an SEIR model is

\begin{equation} \label{eq:intrins-gen-time}
    f(t) = \frac{\kappa\gamma}{\kappa - \gamma}\left(\exp{(-\gamma t)} - \exp{(-\kappa t)}\right)
\end{equation}
(see~\citet{Champredon2018} Table 1). 

\subsubsection{$R_t$ inference using the compartmental model}

We varied $\beta$, $\kappa$ and $\gamma$ during optimisation and inference and fitted our model to prevalence data only. We took the resultant posterior distribution and forward-simulated the model to find means and credible intervals for the population in each of the SEIR model compartments.

\citet{Dankwa2022} found that when fitting an SEIR model (Eq~\ref{eq:seir-model}) to prevalence data, $\beta$ is globally structurally identifiable, meaning that there exists a unique solution in all of parameter space, but $\gamma$ and $\kappa$ are only identifiable in a local region of parameter space. For this reason, in~\nameref{supplementary} Section 2 we verify that this restricted identifiability does not affect the overall conclusions of this work, obtaining equivalent results with fixed $\gamma$ and $\kappa$ at their true values from Epiabm (where they correspond to the inverse mean recovery period and inverse mean incubation period respectively).

In both approaches, we used an order one auto-regressive log-likelihood for inference \citep{lambert2023autocorrelated} --- see~\nameref{supplementary} Section 3 for a description of this log-likelihood. We chose this particular inference approach due to the linear relationship that we observed between the residual errors in the prevalence and the lag~\citep{Hamilton1994} (see~\nameref{supplementary} Section 3). This method also introduces two extra hyperparameters to be inferred: $\rho_{I}$ and $\sigma_{I}$. We followed a multi-step parameter inference routine, first running ten optimisations maximising this log-likelihood (using a Covariance Matrix Adaptation Evolution Strategy (CMA-ES) optimiser~\citep{Hansen2003, Clerx2019Pints}). We then refined these estimates by running 4 chains of 80000 iterations of the Haario-Bardenet Adaptive Covariance Matrix algorithm~\citep{Haario2001, Johnstone2016, Clerx2019Pints} to find posterior distributions for these parameters, starting each chain close to the optimal parameter set obtained from the initial step. The prior distributions used are detailed in ~\nameref{supplementary} Section 4, and the first half of each chain was discarded as a warm-up period. 

Chain convergence was verified by ensuring that $\hat{R}$ values of less than 1.01 were obtained for all parameters. The $\hat{R}$ value accounts for both the intra-chain and inter-chain variability, and is close to 1 when each individual chain is close to a stationary distribution and when the chains are similar to one another \citep{Gelman1992}. All optimisation and inference algorithms described in this subsection were run in Python using the PINTS package~\citep{Clerx2019Pints}.

\subsubsection{Prediction of the case reproduction number}

To find centred 95\% credible intervals for $R_{t}^{\mathrm{case}}$, we first sampled 1000 triples ($\beta$, $\kappa$, $\gamma$) from the posterior distribution, and then forward-simulated the model with these parameters to find the compartmental curves. We then used Eq~\ref{eq:seir-instantaneous-Rt} to find samples for $R_{t}^{\mathrm{inst}}$ and inserted these into Eq~\ref{eq:seir-case-Rt} for $R_{t}^{\mathrm{case}}$. From here, we calculated the 2.5\textsuperscript{th} and 97.5\textsuperscript{th} percentiles of $R_{t}^{\mathrm{case}}$. 

\subsection{Renewal model and inference framework} \label{sec:methods-branch-model}

\subsubsection{$R_t$ inference using the renewal model}

The renewal model~\citep{cori2013new, Thompson2019} was employed to estimate the $R_t^{\mathrm{inst}}$ value using incidence data derived from the agent-based model (see~\nameref{supplementary} Section 5 for more details). In this approach, the number of cases an individual (infected $s$ days ago) causes on day $t$ is drawn from a Poisson distribution with a mean of $R_t^{\mathrm{inst}}w_s$, resulting in a total infection rate of $R_t^{\mathrm{inst}}\sum_{s=1}^{t}I_{t-s}w_s$. This is conditional on the previous incidence values $I_0, I_1, \dots, I_{t-1}$ and generation time distribution $w_s$. Therefore, the incidences are assumed to obey:
\begin{equation} \label{eq:renewal-model}
    I_{t} \sim \mathrm{Poisson}\left(R_t^{\mathrm{inst}}\sum_{s = 1}^{t}w_{s}I_{t - s}\right).
\end{equation}

We set the daily incidence values $I_t$ equal to the number of agents entering the exposed compartment on day $t$. We set the $\{w_s\}$ equal to the generation time distribution observed during the Epiabm simulations, based on the time between infector and infectee entering the exposed compartment.

We chose a gamma distribution as an appropriate conjugate prior and utilised the inference framework implemented through the Python package Branchpro \citep{creswell2022heterogeneity}. We adopted a conservative prior, with a large standard deviation ($\sigma = 5$) and a mean of $5$. Unlike the inference settings selected by
\citep{Thompson2019, creswell2022heterogeneity} (which analyzed outbreaks with far fewer daily cases) we did not use a sliding window to regularize estimates of $R_t^{\mathrm{inst}}$ for the Epiabm datasets.

Following this, we converted the instantaneous reproduction number into a case reproduction number using Eq~\ref{eq:seir-case-Rt}. For the renewal model the intrinsic generation time distribution is $f(t)=w_t$.

\section{Results}

\subsection{Epiabm can simulate a COVID-19-like epidemic in Northern Ireland}

Using the agent-based model introduced in Section \ref{sec:agent-based-model}, we generated synthetic data for an epidemic across Northern Ireland, with a population of 1,837,198 people. The epidemic was initialised with 100 infected individuals in the same local region and simulated for 90 days (chosen to capture extinction of the epidemic).

\begin{figure}
    \centering
    \includegraphics[width=0.8\linewidth]{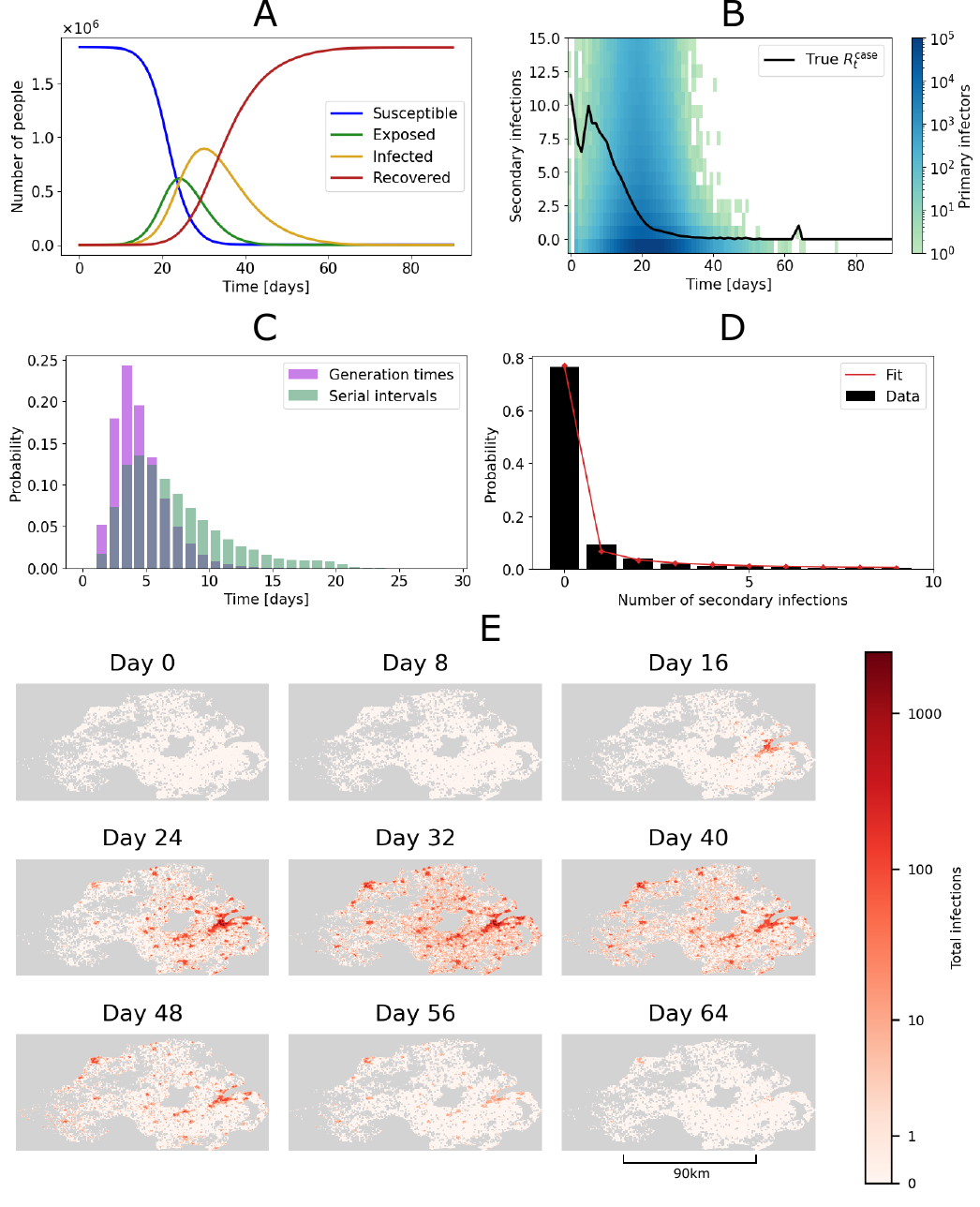}
    \caption{{\bf A simulated epidemic across Northern Ireland using Epiabm with an infection radius of 90km.} A: The SEIR compartmental aggregates for our simulation, starting with 100 infected individuals. B: $R_{t}^{\mathrm{case}}$ and the distributions of secondary infections. Each shading represents the number of primary infectors, whose infections started at a specific time point, that go on to have a specific number of secondary infections over the course of their infection. C: The generation time distribution and serial interval distribution, as parameterised within the Epiabm model. D: Offspring distribution for the Epiabm model, and probability mass function for a negative binomial distribution fit to the offspring distribution data between 0 and 9 secondary infections (10 or more secondary infections are suppressed from the plot for visibility). E: Spatial behaviour of the infection wave over time. Each individual frame represents a snapshot in time, where the colour of each pixel represents the total number of infected individuals in that location. The plot was produced using EpiGeoPop \citep{Ellmen_EpiGeoPop}.}
    \label{fig:synthetic-Epiabm-panel}
\end{figure}

In Fig~\ref{fig:synthetic-Epiabm-panel}A, we plot the aggregated transmission dynamics. The epidemic is seeded on day 0, and prevalence peaks on day 32 before subsiding, and has almost diminished entirely by day 64 (see~\nameref{video:1} for an animation of this simulation). Despite the spatial complexity in our model, and visible in  Fig~\ref{fig:synthetic-Epiabm-panel}E, the SEIR curves exhibit characteristic model epidemic shapes familiar from spatially averaged models.

The values of $R_t^{\mathrm{case}}$ and the variation in the number of secondary infections can be seen in Fig~\ref{fig:synthetic-Epiabm-panel}B. For ease of notation, we will refer to $R_{t}^{\mathrm{case}}$ as $R_{t}$ throughout the results section, but see Section \ref{sec:reproduction-number} for differences between this and the instantaneous $R_{t}$. The vertical axis represents the number of secondary infections per individual and the colour represents the number of individuals that go on to cause that number of secondary infections. This means the superspreaders of the epidemic appear at the top of the plot. These individuals dramatically affect the value of $R_{t}$ near the beginning of the epidemic when few people are infected, and so this causes the true $R_{t}$ curve to fluctuate. Note the delayed nature of this definition of the case reproduction number (see Eq~\ref{eq:seir-case-Rt}); the value of $R_{t}$ on a given day is primarily affected by events in the future, and so $R_{t}$ cannot be recorded accurately until a few days afterwards.

In Fig~\ref{fig:synthetic-Epiabm-panel}C, we plot the generation time and serial interval distributions, as extracted from Epiabm for this epidemic simulation \textemdash\ these are the same for both simulations presented in this paper. Fig~\ref{fig:synthetic-Epiabm-panel}D displays the offspring distribution and its fit to a negative binomial distribution. The model provides a relatively good fit, suggesting that the Epiabm simulation is realistic in this regard. 

\subsection{Compartmental and renewal models can both infer $R_{t}$ from data with low spatial heterogeneity} \label{sec:inference-comparison-r1}

\begin{figure}
    \centering
    \includegraphics[width=0.9\linewidth]{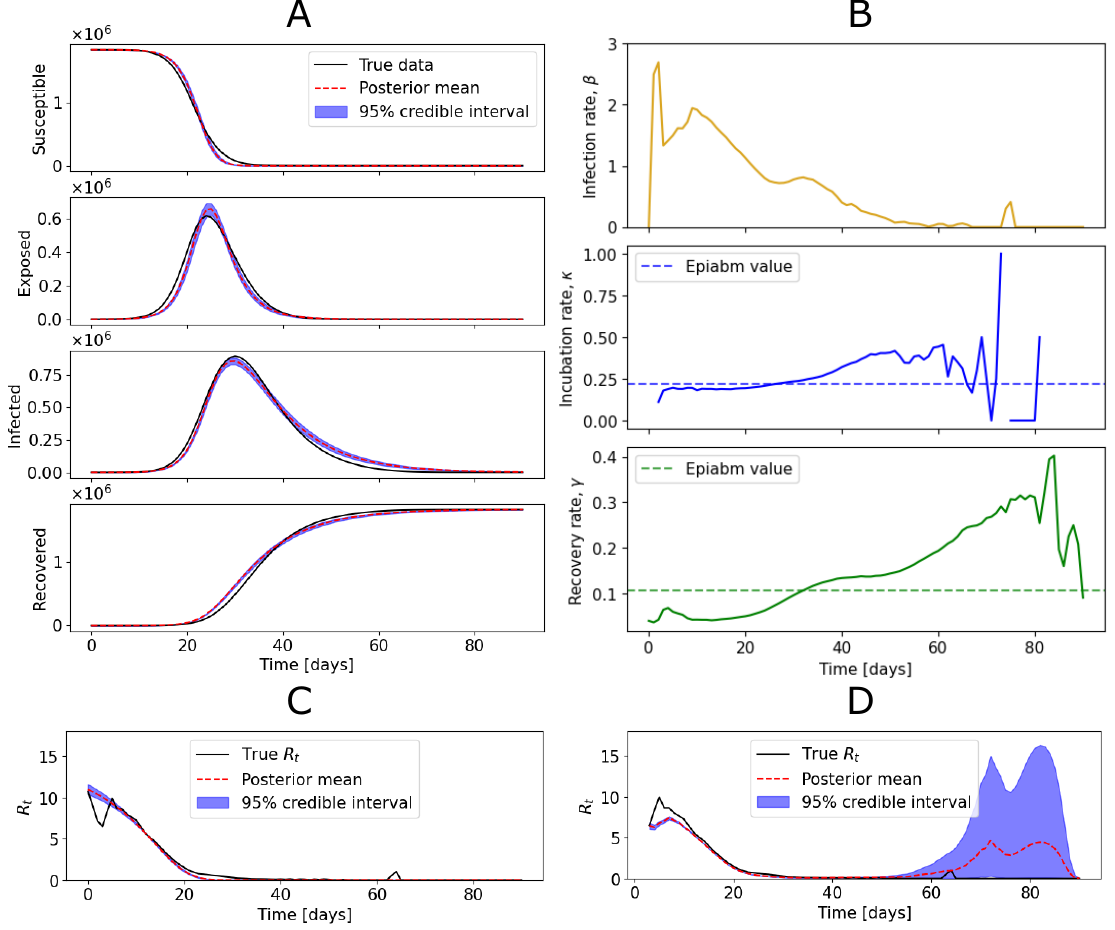}
    \caption{{\bf SEIR and renewal model inference results from the dataset with an infection radius of 90km.} A: Inferred time series for the SEIR compartmental aggregates with 95\% credible intervals. We sampled 1000 triples ($\beta$, $\kappa$, $\gamma$) from the posterior distribution, simulated the model and calculated the mean, 2.5\textsuperscript{th} percentile and 97.5\textsuperscript{th} percentile to generate the plots. B: The time-varying estimates for the SEIR parameters $\beta$, $\gamma$ and $\kappa$. These are calculated from Eq~\ref{eq:seir-model} using the synthetic compartmental data curves and their derivatives, and compared to true values extracted from Epiabm (see~\nameref{supplementary} Section 6). C: Posterior mean $R_{t}$ curve with 95\% credible intervals for the compartmental model. Using the 1000 triples found in A, we used Eq~\ref{eq:seir-case-Rt} and Eq~\ref{eq:intrins-gen-time} to find $R_{t}$ (using the forward-simulated susceptible curve). We then found the mean and 95\% credible intervals to generate the plots. D: Posterior mean $R_{t}$ curve with 95\% credible intervals for the renewal model.}
    \label{fig:seir-inference-panel}
\end{figure}

Here, we compare methods for inferring $R_{t}$ from synthetic data using two conventional epidemiological models: the SEIR model and a renewal model. We sampled 1000 parameter sets ($\beta$, $\kappa$, $\gamma$) from our posterior and forward-simulated the model to examine fits of the compartmental model to the prevalence data. Although the data originate from a complex agent-based model with spatial and age effects, the SEIR model fits the aggregate data closely; however, the synthetic curves do not always lie within the 95\% predictive interval (Fig~\ref{fig:seir-inference-panel}A).

We attribute this to inherent temporal variation in $\beta$, $\kappa$ and $\gamma$, which is not captured by the conventional SEIR model (Eq~\ref{eq:seir-model}). We extracted time-varying expressions for $\beta$, $\kappa$ and $\gamma$ by rearranging Eq~\ref{eq:seir-model}, and computed crude estimates for the time-varying model transition rates observed in synthetic data by evaluating the piecewise gradient of the compartment populations. For a description of this method, along with a validation on SEIR-generated data, see~\nameref{supplementary} Section 6. Despite the visual agreement between the inferred and true compartment populations in Fig~\ref{fig:seir-inference-panel}A, there is substantial variation in the time-varying estimates of the epidemiological parameters in Fig~\ref{fig:seir-inference-panel}B, suggesting that, in actuality, the transmission dynamics varied over time. This hints that the SEIR model with constant parameters is fundamentally misspecified for this complex synthetic data.

We inserted our 1000 ($\beta$, $\gamma$, $S(t)$) MCMC samples in Eq~\ref{eq:seir-instantaneous-Rt} to produce a posterior distribution for $R_{t}^{\mathrm{inst}}$ and fed the ($\gamma$, $\kappa$, $R_{t}^{\mathrm{inst}}$) samples into Eq~\ref{eq:seir-case-Rt} and Eq~\ref{eq:intrins-gen-time} to get a posterior distribution for $R_{t}^{\mathrm{case}}$. We then found the posterior mean and 95\% credible interval for the $R_{t}$ curves (Fig~\ref{fig:seir-inference-panel}C), which fit well to the true underlying $R_{t}^{\mathrm{case}}$ from Epiabm.

We contrast these results with the $R_t$ values estimated from the renewal model. Fig~\ref{fig:seir-inference-panel}D shows a close alignment between the estimated $R_t$ and the true $R_t$ values derived from the Epiabm simulation, including the observed fluctuations in the true $R_t$.

Despite this close alignment, it is worth highlighting that our posterior credible interval for $R_{t}$ does not fully overlap with its true value, with consistent underestimation at early times of the simulation. This model misspecification is to be expected; the formulation of uncertainty within the renewal model assumes data generated from a renewal model framework, and only captures the inherent stochasticity in renewal model dynamics. Data derived from the agent-based model (similarly to real-world data) violates the underlying assumptions of this \textemdash\ for example age-dependent variation in infection dynamics across the population introduces heterogeneity in the generation time between individuals. Previous work has demonstrated the validity of inference approaches based on the renewal model, when the simulated data is also generated from a simplified renewal model that does satisfy these assumptions \citep{Gostic2020}.

\subsection{Renewal models infer $R_{t}$ more accurately than compartmental models for spatially heterogeneous data}

The data presented thus far were obtained using a large infection radius ($r = 90$ km); a parameter defined in Epiabm as the furthest distance over which an individual can carry out an infection. When we reduce this radius to $18$km, the wave of infection takes longer to travel around the country and tends to be more spatially heterogeneous. Note that in Epiabm this parameter only affects the choice of potential infectees in each infection event, and the serial interval, generation time and offspring distributions are approximately unchanged between the two simulation cases.

\begin{figure}
    \centering
    \includegraphics[width=0.8\linewidth]{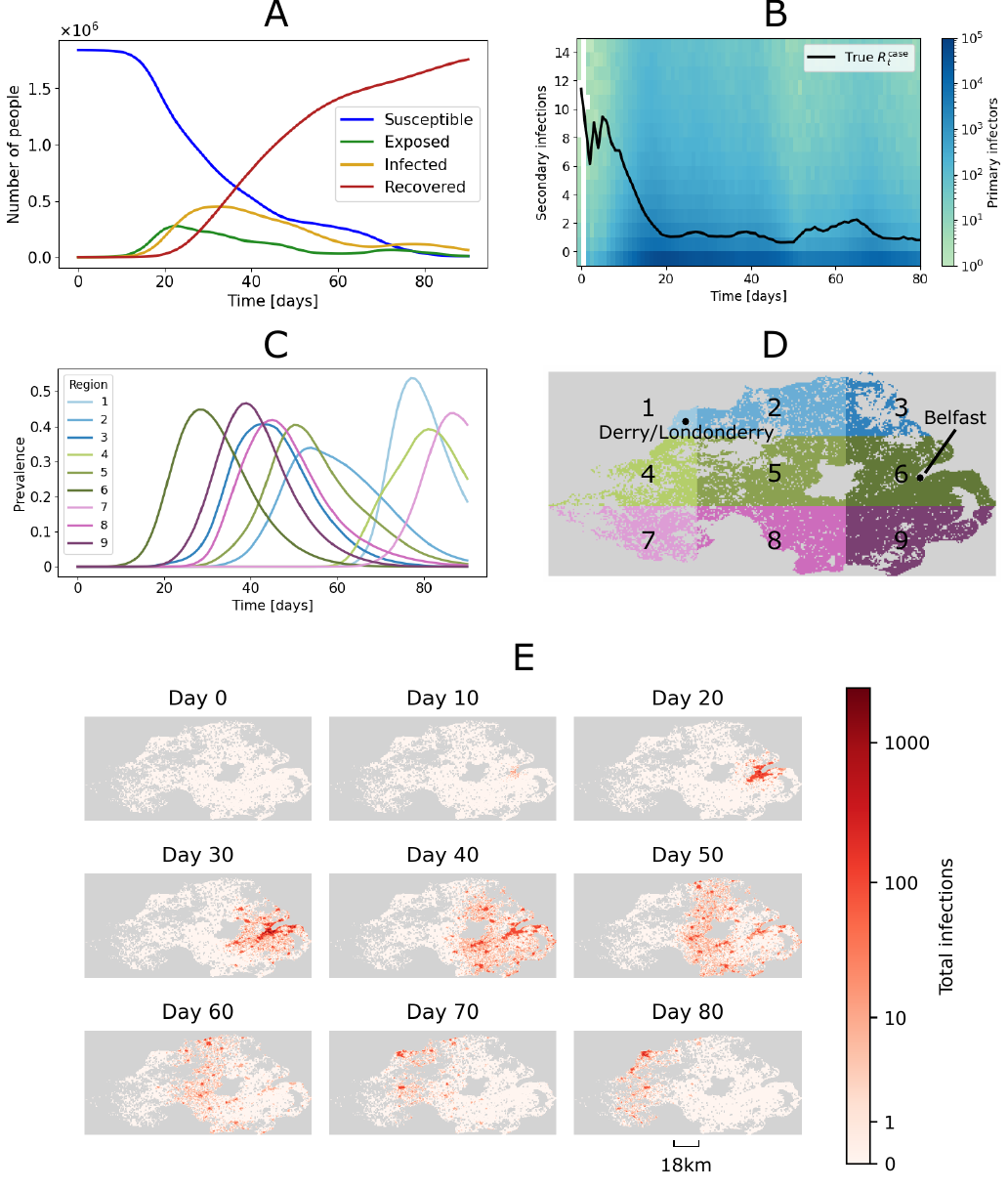}
    \caption{{\bf A simulated epidemic across Northern Ireland using Epiabm with an infection radius of 18km.} A: The SEIR compartmental traces for our simulation, starting with 100 infected individuals --- note the multiple peaks in the infected compartment. B: $R_{t}^{\mathrm{case}}$ and the distributions of secondary infections over time. Each shading represents the number of primary infectors, whose infections started at a specific time point, that go on to have a specific number of secondary infections over the course of their infection. C: Prevalence profiles in different spatial regions of Northern Ireland. D: The geographical locations of the nine spatial regions in panel C. E: Spatial behaviour of the infection wave over time. Each individual frame represents a snapshot in time, where the colour of each pixel represents the total number of infected individuals in that location. Panels D and E were produced using EpiGeoPop \citep{Ellmen_EpiGeoPop}.}
    \label{fig:synthetic-Epiabm-panel-02}
\end{figure}

Fig~\ref{fig:synthetic-Epiabm-panel-02}C, shows that the infection starts near Belfast (Region 6 of Fig~\ref{fig:synthetic-Epiabm-panel-02}D). From here, the wave of infection travels from East to West (darker to lighter colours in Fig~\ref{fig:synthetic-Epiabm-panel-02}C) and reaches Derry/Londonderry (Region 1) at a much later time; this large population centre contributes to the secondary infection peak seen in Fig~\ref{fig:synthetic-Epiabm-panel-02}A, with a corresponding peak in the true $R_{t}$ shown in Fig~\ref{fig:synthetic-Epiabm-panel-02}B. Supplementary~\nameref{video:2} provides an animation of this simulation, with corresponding snapshots presented in Fig~\ref{fig:synthetic-Epiabm-panel-02}E. We now compare the compartmental and renewal model inference frameworks when fitting to these data.

\begin{figure}
    \centering
    \includegraphics[width=0.9\linewidth]{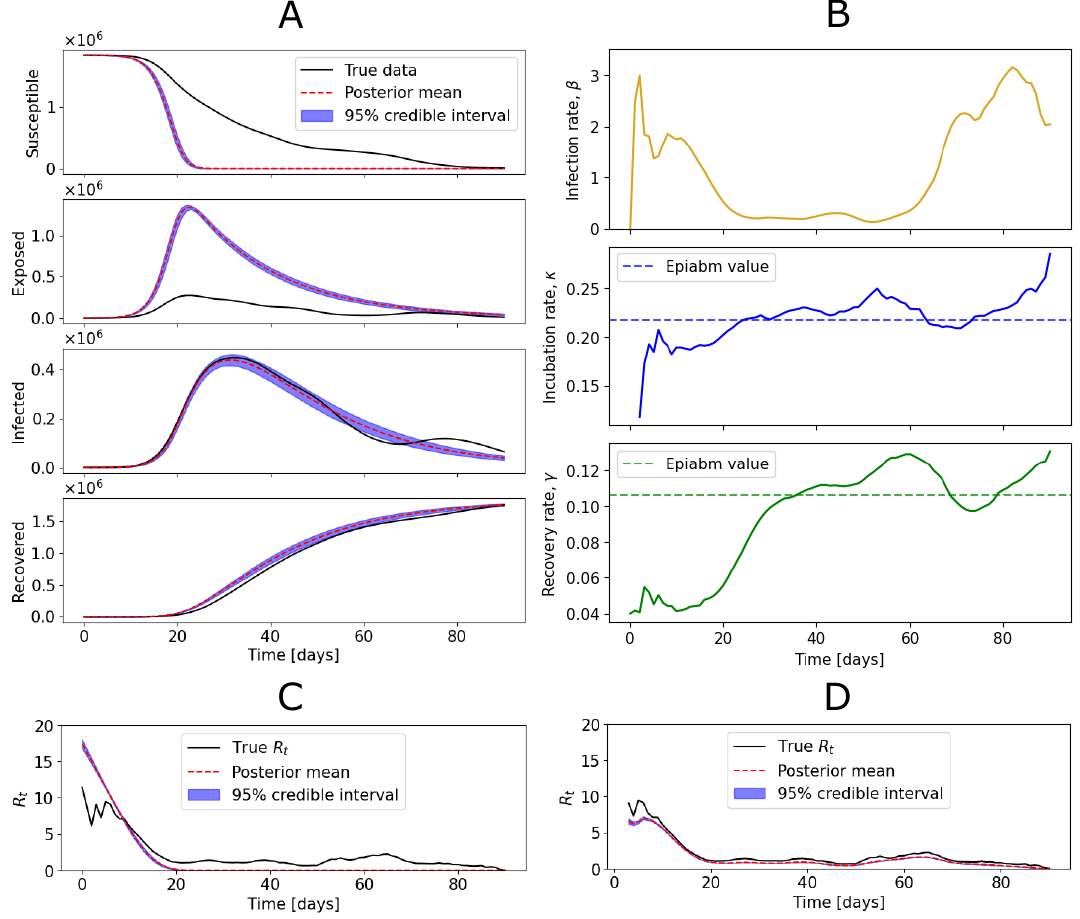}
    \caption{{\bf SEIR and renewal model inference results from the dataset with an infection radius of 18km.} A: Inferred time series for the SEIR compartments with 95\% credible intervals for the compartmental model. We sampled 1000 triples ($\beta$, $\kappa$, $\gamma$) from the posterior distribution and simulated the model. From here we calculated the mean, 2.5\textsuperscript{th} percentile and 97.5\textsuperscript{th} percentile to generate the plots. B: The time-varying estimates for the SEIR parameters $\beta$, $\gamma$ and $\kappa$. These are calculated from Eq~\ref{eq:seir-model} using the synthetic compartmental data curves and their derivatives (see \nameref{supplementary} Section 6). C: Posterior mean $R_{t}$ curve with 95\% credible interval. Using the 1000 triples found in A, we used the formulae in Eq~\ref{eq:seir-case-Rt} and Eq~\ref{eq:intrins-gen-time} to find $R_{t}$ (using the forward-simulated susceptible curve each time), and plotted the mean and 95\% credible interval. D: Posterior mean $R_{t}$ curve and 95\% credible interval for the renewal model.}
    \label{fig:rsmall-panel}
\end{figure}

We first followed the same approach as in Section \ref{sec:inference-comparison-r1} to infer $R_{t}$ using the compartmental model, and Fig~\ref{fig:rsmall-panel}A shows the means and credible intervals for the SEIR compartment populations over time. Due to mechanisms in the underlying ABM that cannot be captured by the simple SEIR model, the fit to the true $R_{t}$ curve is poor. This remains true even when we fit our model to all four compartmental curves and not just the prevalence (\nameref{supplementary} Section 7). The time-varying estimates for the epidemiological parameters (Fig~\ref{fig:rsmall-panel}B) change considerably over time. In addition, the secondary peak in infections (Fig~\ref{fig:rsmall-panel}A) is mirrored by a secondary peak in the transmission parameter, $\beta$. Fig~\ref{fig:rsmall-panel}C shows that our inferred $R_{t}$, calculated in the same way as before, initially overestimates the true $R_{t}$ while subsequently failing to capture the later peaks in $R_{t}$. This further demonstrates that the inference framework based on the SEIR model is unsuitable for this dataset.

Fig~\ref{fig:rsmall-panel}D shows that the renewal model successfully estimates the dynamics of $R_t$ with a small uncertainty represented by the width of the credible interval. While this model suffers from the same formal misspecification limitations, as discussed in Section \ref{sec:inference-comparison-r1}, the overall $R_{t}$ dynamics are captured more faithfully than they are for the simulation with a larger infection radius. The renewal model is also able to capture secondary infection waves that occur later in the simulation, and is generally sensitive to small fluctuations of $R_t$. Despite this, it is also worth noting that the estimate of $R_{t}$ is highly sensitive to the generation time distribution used, as demonstrated in~\nameref{supplementary} Section 8.

\section{Discussion}
In this study, we presented tools to simulate and sample from an epidemiological agent-based model to enable benchmarking of different $R_{t}$ inference tools against ground-truth data with age and spatial heterogeneity. The complexity of this model replicated many aspects of real-world data (such as demographic or spatial heterogeneity) that confound basic inference methods. Furthermore, our use of synthetic data allowed direct comparison of the inferred $R_{t}$ time series with a known ground truth extracted from the model.

We compared two standard frameworks (based on an SEIR compartmental model and a renewal model) for inferring $R_{t}$ traces from synthetically generated data for an outbreak in Northern Ireland. Both of these models were able to infer $R_{t}$ accurately when there was minimal spatial heterogeneity in the epidemic simulation, but the renewal model was better able to capture complex infection dynamics (such as multiple peaks in incidence) from simulations with greater spatial heterogeneity. The poorer performance of the compartmental model may be attributed to its inability to capture the spatial heterogeneities in the ground-truth data.


Many extensions to the inference approaches considered here have been developed. Compartmental models may be extended to better describe the true processes which spread the disease, through the introduction of additional compartments. These can be used to represent different ages, locations, vaccination statuses, and other relevant individual characteristics~\citep{Keeling2022, van2022learning}. Alternatively, time variation may be introduced in the transition parameters in SEIR models \citep{Chen2020, Hong2020, girardi2023seir, zelenkov2023analysis}, allowing the model to fit more complex infection curves, such as those resulting from spatial heterogeneity in the regions where the disease is spreading. However, these approaches introduce large numbers of unknown parameters, increasing the complexity and computational cost of the associated inference frameworks.

The Epiabm datasets used in this paper involved up to hundreds of thousands of cases per day, representative of a disease which rapidly infects a substantial portion of the population. However, many real diseases are less prevalent, and when applying the methods considered in this paper to datasets with fewer cases, greater uncertainty in model fits may be expected. To increase the precision of $R_t$ estimates in the renewal equation model in these settings, the sliding window heuristic algorithm \citep{Thompson2019} would have to be used.

Alternative approaches to modelling time variation in the reproduction number in renewal models include Bayesian smoothing and Gaussian processes \citep{ abbott2020estimating, parag2021improved}. Another possibility is to assume that daily case numbers follow a negative binomial distribution \citep{ho2023accounting} (rather than the more standard Poisson distribution we employ). Given that we observed overly precise posterior uncertainty estimates of $R_t$ in many instances, further work to relax the assumptions of the underlying case distribution would support the application of the renewal approach on realistic data, with more realistic uncertainty estimates.

Agent-based models such as Epiabm are also stochastic, with different simulations at the same parameter values leading to different epidemic trajectories. A practical limitation of our study is that we applied our inference methods to just two sets of data, rather than averaging the expected performance of the inference approaches over many realizations of Epiabm simulations, for the two cases. Due to the runtime of complex agent-based models, incorporating uncertainty in model simulations in a comprehensive way would require significant advances in computational budget, or the development of faster implementations/approximations of the ABM simulations.

Our approach to implementing the renewal model inference procedure used in this paper was rooted in a number of simplifying assumptions. To approximate the generation interval for use in the renewal equation, we simply extracted and supplied the empirically observed intervals from the Epiabm simulation, without performing any corrections for the biases which may cause realized generation intervals to differ from the intrinsic value of the generation interval distribution \citep{champredon2015intrinsic}. Given the sensitivity of renewal model inference results to generation interval assumptions observed in this paper, further work to implement and evaluate robust methods for learning generation intervals which are valid for inference, e.g., \citep{Park2020}, would be highly valuable. Furthermore, our comparisons between ground truth and inference results focused on the case $R_t$, whereas epidemiological inference approaches have typically been benchmarked with respect to instantaneous reproduction numbers, which are directly outputted by the renewal model inference approach. The case reproduction number provides an attractive, interpretable summary statistic of real-world (or ABM-generated) transmission data, but further work may be necessary to investigate the robustness of methods for converting instantaneous to case reproduction number inference results, given their dependence on knowledge of the generation interval.

Throughout this study, we provided our inference methodologies with perfect knowledge of the disease prevalence or incidence as extracted from the agent-based model simulations. In real diseases, however, only an incomplete picture is available, informed by disease surveillance strategies such as the reporting of symptomatic cases, testing and reporting of at-risk individuals, and prevalence surveys of randomly selected subsets of the population, e.g.\ \citep{chadeau2022sars}. Similarly, when cases or deaths are available, information about their values is inevitably affected by reporting delays and under reporting, which can affect the ability to learn $R_t$ \citep{parag2022quantifying}; for some diseases, observed time series are also corrupted by pronounced weekly periodicities \citep{gallagher2023identification}. Although we do not present an investigation of these processes and their impact on parameter inference in this paper, the simulation and inference framework we have adopted for our experiments can be straightforwardly extended to supply the inference methods with more realistic, noisy or biased estimates of epidemiological observations to investigate the effects on their ability to recover true parameter values. Extensible open-source software, as emphasized and made available in this study, makes it straightforward for such features to be added on top of the existing functionality, without requiring researchers to invest in reimplementing the underlying agent-based simulator or inference algorithms.

In conclusion, we found that using an agent-based model to generate a ground-truth $R_{t}$ provides a useful benchmark on which to compare various inference methods for $R_{t}$. In this context, we found a renewal model was able to capture the complex infection dynamics that arose from spatial heterogeneity more accurately than a compartmental model, however, this may not apply to all estimation problems or disease contexts. Crucially, the model used to generate the synthetic data was different to the models used for the inference of $R_{t}$, demonstrating the capability of this approach to mimic validation on real-world data with the additional benefit of having a ground truth. This method may be further applied to benchmark other inference models beyond those considered in this work.

Despite our attention to $R_t$, the analysis we perform in this paper could also be straightforwardly extended to consider other sensible parameters, such as growth rates \citep{parag2022epidemic}, or to direct predictions of future cases. Further investigation and promulgation of summary statistics such as growth rates (which can be computed without direct knowledge of the generation interval) may be a particularly productive direction, given the challenges and complexities involved in learning generation intervals, and the sensitivity of our renewal model results to generation interval values. We hope that our work will promote further benchmarking, comparison and cross-validation of different inference methods using complex synthetic data when estimating a range of epidemiological parameters.

\section*{Data Accessibility}
We used the Epiabm package to produce all synthetic data for this study. See \url{https://github.com/SABS-R3-Epidemiology/epiabm/tree/main/python_examples/NI_example} for further details on parameter values and configuration of the simulation. The workflows for conducting all model inference are contained within the following directory: \url{https://github.com/SABS-R3-Epidemiology/seirmo/tree/main/examples/epiabm_rt_inference/northern_ireland}. 

\section*{Glossary}

\noindent Agent-based model: An algorithm that explicitly models
people as individual agents with distinct characteristics and behaviour.

\noindent Case reproduction number, $R_{t}^{\mathrm{case}}$: The time-dependent average number of secondary cases each person infected with the disease will go on to cause, whereby each secondary case is associated with the time of entry into the exposed compartment.

\noindent Compartmental model: In the context of epidemiology, a coupled system of differential equations tracking the evolution of the number of individuals in each infection compartment (for example, Susceptible, Exposed, Infected and Recovered) over time.

\noindent Generation time distribution: The distribution of the times between infection of a primary and secondary case.

\noindent Instantaneous reproduction number, $R_{t}^{\mathrm{inst}}$: The time-dependent average
number of secondary cases each person infected with the disease will go on
to cause, whereby each secondary case is associated with the time of the infection event.

\noindent Renewal model: A model of disease incidence over an epidemic, in which the current incidence is related to the instantaneous reproduction number, generation time and all previous incidences.

\noindent Serial interval distribution: The distribution of the times between symptom onset of a primary and secondary case.

\section*{Acknowledgements}
The authors wish to acknowledge the work of Antonio Mastromarino, on software that contributed to the work in this paper.

\section*{Authorship Contribution Statement}

Matthew Ghosh: Data curation, Investigation, Methodology, Software, Visualization, Writing – original draft, Writing – review \& editing

Yunli Qi: Data curation, Investigation, Methodology, Software, Visualization, Writing – original draft, Writing – review \& editing

Abbie Evans: Data curation, Software, Visualization, Writing – review \& editing

Thomas Reed: Data curation, Software, Writing – review \& editing

Lara Herriott: Supervision, Writing – review \& editing

Ioana Bouros: Software, Supervision, Writing – review \& editing

Ben Lambert: Conceptualization, Supervision, Writing – review \& editing

David Gavaghan: Conceptualization, Funding acquisition, Supervision, Writing – review \& editing

Katherine Shepherd: Investigation, Methodology, Supervision, Visualization, Writing – original draft, Writing – review \& editing

Richard Creswell: Conceptualization, Methodology, Supervision, Visualization, Writing – original draft, Writing – review \& editing

Kit Gallagher: Conceptualization, Methodology, Software, Supervision, Visualization, Writing – original draft, Writing – review \& editing

\section*{Declaration of Competing Interest}
The authors declare no competing interests.

\section*{Funding}
All authors acknowledge funding from the EPSRC CDT in Sustainable Approaches to Biomedical Science: Responsible and Reproducible Research - SABS:R3 (EP/S024093/1). 

\bibliographystyle{elsarticle-harv} 
\bibliography{bibliography}

\section*{Supplementary Material}\label{supplementary}

\subsection*{1 - Key parameter values for the simulation of Northern Ireland} \label{sec:NI-params}
In this section, we present a summary of important parameter values used in the agent-based simulation of Northern Ireland (see Table~\ref{table:NI-params}). The two simulations presented in the main text only differ by their infection radius, $r$, which is 1.0 degrees (approximately 90 km) in the first simulation and 0.2 degrees (approximately 18 km) in the second. 

\begin{table}[!ht]
\centering
\caption{
{\bf Key parameter values for the simulation of Northern Ireland}}
\hspace*{-1.5cm}\begin{tabular}{|l|l|l|l|}
\hline
Parameter & Description & Value & Unit\\
\hline
Infection radius & Maximum distance an individual can travel & 1.0 or 0.2 & degrees\\
\hline
Latent period & Mean duration in the exposed compartment & 4.59 & days\\
\hline
Infection period & Mean period from exposure to recovery & 14 & days\\
\hline
Simulation end time & Length of simulation & 90 & days\\
\hline
Initial infected number & Number of infected individuals initially & 100 & people\\
\hline
\end{tabular}
\begin{flushleft}  We note that the inverse of the latent period is an estimate for $\kappa$ and that the inverse of the infectious period minus the latent period is an estimate for $\gamma$. See \url{https://github.com/SABS-R3-Epidemiology/Epiabm/tree/main/python_examples/NI_example} for a full list of parameter values and further details on the simulation set-up, as well as complete simulation code. 
\end{flushleft} \label{table:NI-params}
\end{table}

\subsection*{2 - Parameter identifiability in the compartmental model}

As discussed in Section 2.2.3 of the main text, when inferring $\beta$, $\kappa$, and $\gamma$ from a standard SEIR model, only $\beta$ is globally structurally identifiable; $\kappa$ and $\gamma$ are structurally identifiable only in a local region of parameter space~\citep{Dankwa2022}. Hence, in this section, we fix $\kappa$ and $\gamma$ at their means from Epiabm (see Table~\ref{table:NI-params}) and infer $\beta$, $\rho_{I}$ and $\sigma_{I}$ (as defined in Section 3).

\begin{figure}
    \centering
    \includegraphics[width=1.0\linewidth]{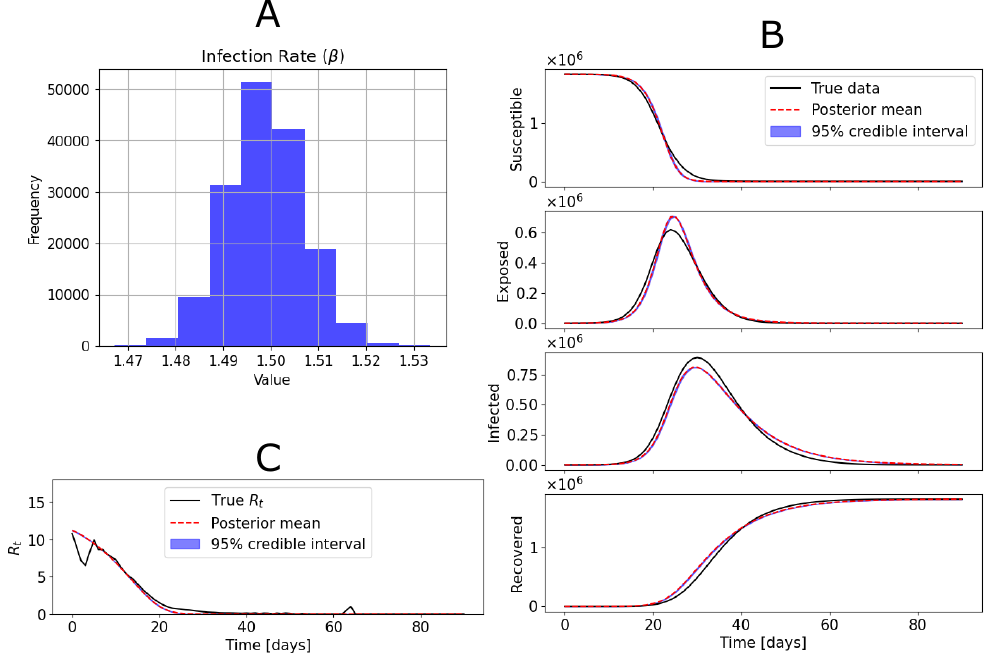}
    \caption{{\bf Results from inferring $\beta$ only in the compartmental ODE model.} We simulated 4 chains for 80000 iterations and discarded the first half as a warm-up period. A: Posterior samples for $\beta$ when fitting the SEIR model to prevalence data with infection radius $1$, fixing $\gamma = 1 / 9.41$, $\kappa = 1 / 4.59$, the mean values from Epiabm (see Table~\ref{table:NI-params}). B: Inferred time series for the SEIR compartments with 95\% credible intervals. We sampled 1000 values of $\beta$ from the posterior distribution and simulated the model. From here we calculated the mean, 2.5\textsuperscript{th} percentile and 97.5\textsuperscript{th} percentile to generate the plots. C: Inferred $R_{t}$ curve with 95\% credible intervals. Using the 1000 triples found in A, we used the formula in Eq 3 of the main text to find $R_{t}$ (using the forward-simulated susceptible curve each time). We then found the mean and 95\% credible interval to generate the plots.}
    \label{fig:inferring-beta}
\end{figure}

Following the same optimisation and inference procedures as described in Section 2.2.3 of the main text, we achieved $\hat{R}$ values of less than 1.01 in all cases, and then sampled 1000 values of $\beta$ from our posterior distribution (see Fig~\ref{fig:inferring-beta}A for this distribution). We see from the $R_{t}$ estimate in this case (Fig~\ref{fig:inferring-beta}C) that the magnitude is similar to Fig 2C of the main text, but with a narrower credible interval. The reduced width of the credible interval is due to the fact that $\beta$ is globally structurally identifiable~\citep{Dankwa2022} and so is determined with higher precision.

Note that the constant values of $\gamma$ and $\kappa$ taken from Epiabm are given by the mean of their respective probability distributions, however their time-varying estimates (main text Fig 2B) demonstrate that they vary significantly over time. Despite this, we were able to reproduce the compartmental and $R_{t}$ curves relatively accurately, as the SEIR curves for the Epiabm data followed similar forms to curves generated from the ODE model with constant parameters.

\subsection*{3 - Autocorrelations of prevalence residuals in the compartmental model}

We firstly give a brief description of our order one auto-regressive log-likelihood (see~\citet{Clerx2019Pints}). For each time $t$ and a given parameter vector $\bm{\theta}$, we define the residual $\epsilon_{t}$ as

\begin{equation} \label{eq:residual-defn}
    \epsilon_{t} := x_{t} - f_{t}(\bm{\theta}),
\end{equation}

\noindent where $x_{t}$ is the observed datapoint and $f_{t}(\bm{\theta})$ is the predicted value given a parameter vector. In an order one auto-regressive log-likelihood, the residuals are assumed to obey the following relation:

\begin{equation} \label{eq:residual-AR1}
    \epsilon_{t} = \rho\epsilon_{t - 1} + \nu_{t}.
\end{equation}

The autocorrelation parameter $\rho$ satisfies $0 < \rho < 1$, and the random variables $\nu_{t}$ are i.i.d. $\sim \mathcal{N}(0, \sigma^{2}(1 - \rho^{2}))$. Therefore, each residual is linearly dependent on the previous one, and we can use $\nu_{t}$ to calculate the form of our log-likelihood.

\begin{figure}
    \centering
    \includegraphics[width=1.0\linewidth]{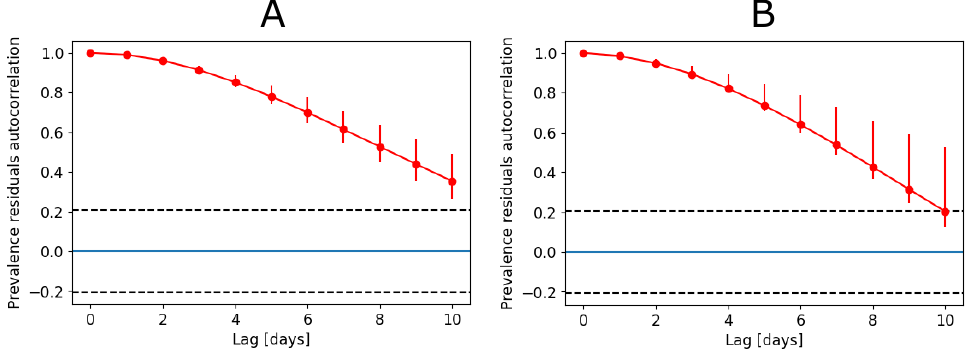}
    \caption{{\bf Prevalence residual autocorrelations for the simulations with large ($r = 1$) and small ($r = 0.2$) infection radii.} The dotted black line denotes the 95\% confidence bounds under the assumption of IID residuals, with the simulated data (red) lying significantly outside of this. A: Infection radius of 1. B: Infection radius of 0.2. Plots produced using PINTS~\citep{Clerx2019Pints}.}
    \label{fig:autocorrelations}
\end{figure}

In order to find the most appropriate form of our log-likelihood function, we plotted the autocorrelations of the prevalence residuals as a diagnostic test. Looking at Fig~\ref{fig:autocorrelations}, we see that when the time lag between residuals is larger than 1 day, the autocorrelation of these residuals in our data have an approximately linear relation with the lag. This is a feature of order one auto-regressive noise as predicted by Eq~\ref{eq:residual-AR1} (see \citet{Hamilton1994} for a description of this), supporting our use of the order one auto-regressive log-likelihood. In contrast, if a Gaussian process underpinned the noise in our data, we would expect the autocorrelated residuals to lie between the two black dashed lines (confidence bounds for the sample autocorrelations under the assumption of IID residuals~\citep{Clerx2019Pints}), as there should be a low or no correlation between residuals generated from a Gaussian noise process.

For our analysis, as we fit the SEIR ODE model to prevalence data from Epiabm (Infected compartment), our log-likelihood function must include an autocorrelation parameter $\rho_{I}$ and a standard deviation $\sigma_{I}$. The subscript `$I$' denotes the Infected compartment.

\subsection*{4 - Prior and posterior distributions of epidemiological parameters}

\begin{figure}
    \centering
    \includegraphics[width=1.0\linewidth]{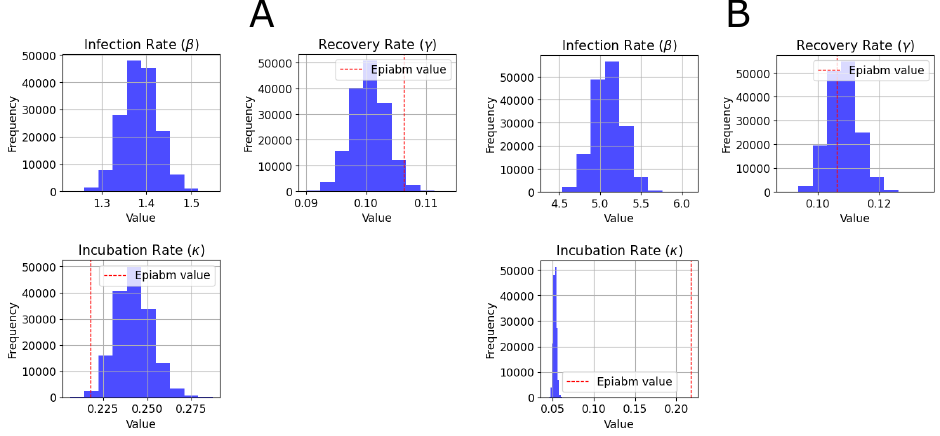}
    \caption{{\bf Posterior samples for the large ($r = 1$) and small ($r = 0.2$) infection radii simulations.} In both cases, we simulated 4 chains for 80000 iterations and discarded the first half as a warm-up period. A: Samples from the marginal distributions for epidemiological parameters $\beta$, $\gamma$ and $\kappa$ when fitting the SEIR model to prevalence data with infection radius $1$. Note that we have marked the mean values of $\gamma$ and $\kappa$ from Epiabm as a point of comparison, while $\beta$ is not parameterised directly in the Epiabm model. B: Samples from the marginal distributions for epidemiological parameters $\beta$, $\gamma$ and $\kappa$ when fitting the SEIR model to prevalence data with infection radius 0.2.}
    \label{fig:posteriors}
\end{figure}

In Section 3.2 of the main text, we presented inferred compartmental and $R_{t}$ curves produced from inferring governing epidemiological parameters $\beta$, $\gamma$ and $\kappa$ from the SEIR ODE model. In Fig~\ref{fig:posteriors} we present samples from the posterior distributions of the governing parameters for the large and small infection radii.

We used the following prior distributions for our parameters:

\begin{align*}
    \beta \sim \mathrm{Uniform}(0.01, 10), \qquad \kappa \sim \mathrm{Uniform}(0.01, 1), \\ 
    \gamma \sim \mathrm{Uniform}(0.01, 1), \qquad \rho_{I} \sim \mathrm{Uniform}(0, 1),
\end{align*}

\noindent For $\sigma_{I}$, we used a Gaussian prior distribution with a mean of 0 and a standard deviation of 1. We also truncated this distribution at 0 so that it only had positive values.

Furthermore, we noted a negative correlation between $\beta$ and $\kappa$ in both sets of inference, and no significant correlation between the other pairs of epidemiological parameters. A correlation of some form was to be expected, as only $\beta$ is globally structurally identifiable from this system~\citep{Dankwa2022}. We addressed this in Section 2, and see our Jupyter notebook for more details on the correlations (\url{https://github.com/SABS-R3-Epidemiology/seirmo/blob/main/examples/epiabm_rt_inference/northern_ireland/northern_ireland_rt_inference.ipynb}).

\subsection*{5 - Details of the renewal model inference approach}

The subsequent discussion provides a summary of previous work by \citet{cori2013new}, replicated for the purposes of the paper. The number of cases a single individual (infected $s$ days ago) causes on day $t$ is drawn from a Poisson distribution with a mean of $R_t^{\mathrm{inst}}w_s$, where $\{w_s\}$ represents the generation time distribution. We denote the incidence at time $t$ by $I_t$.
$$P\left(I_t|I_0,...,I_{t-1},w,R_t\right)=\frac{(R_t\Lambda_t)^{I_t}e^{-R_t\Lambda_t}}{I_t!},$$
with
$$\Lambda_t=\sum_{s=1}^{t}I_{t-s}w_s.$$
With the assumption that $R_t$ is constant over the period $[t-\tau+1,t]$, we can obtain
$$P\left(I_{t-\tau+1},...,I_t|I_0,...,I_{t-1},w,R_{t,\tau}\right)=\prod_{s=t-\tau+1}^{t}\frac{(R_{t,\tau}\Lambda_s)^{I_s}e^{-R_{t,\tau}\Lambda_s}}{I_s!}$$

By using the Bayes' formula with a gamma prior for $R_t$, we find

\begin{align}
P\left( R_{t,\tau} \mid I_0, \dots, I_{t-\tau}, I_{t-\tau+1}, \dots, I_t, w \right) &\propto P\left(I_{t-\tau+1}, \dots, I_t \mid I_0, \dots, I_{t-\tau}, w, R_{t,\tau}\right) P\left(R_{t,\tau}\right)\nonumber\\
&\propto \left( \prod_{s=t-\tau+1}^{t} \frac{\left(R_{t,\tau} \Lambda_s\right)^{I_s} e^{-R_{t,\tau} \Lambda_s}}{I_s!} \right) \left( \frac{R_{t,\tau}^{a-1} e^{-R_{t,\tau}/b}}{\Gamma(a) b^a} \right)\nonumber\\
&\propto R_{t,\tau}^{a + \sum_{s=t-\tau+1}^{t} I_s - 1} e^{-R_{t,\tau} \left(\sum_{s=t-\tau+1}^{t} \Lambda_s + \frac{1}{b}\right)} \prod_{s=t-\tau+1}^{t} \frac{\Lambda_s^{I_s}}{I_s!} \frac{1}{\Gamma(a) b^a}\nonumber\\
&\propto R_{t,\tau}^{a + \sum_{s=t-\tau+1}^{t} I_s - 1} e^{-R_{t,\tau} \left( \sum_{s=t-\tau+1}^{t} \Lambda_s + \frac{1}{b} \right)} \prod_{s=t-\tau+1}^{t} \frac{\Lambda_s^{I_s}}{I_s!} \nonumber
\end{align}

The resulting posterior distribution of $R_t$ follows a gamma distribution. To construct an uninformative prior, the parameters are set to $a = 1$ and $b = 5$, which results in a large standard deviation for the prior distribution. This choice allows the prior to exert minimal influence on the posterior, thereby letting the data primarily determine the inferred posterior.

\subsection*{6 - Validation of use of crude time-varying parameter estimates}

We hypothesised in Section 3.2 of the main text that the true epidemiological parameters ($\beta$, $\kappa$ and $\gamma$) vary substantially over time, using crude time-varying estimates as our evidence for this. In this section, we detail this procedure, along with a validation of the method on data generated from an SEIR compartmental model (\ref{eq:seir-model}).

Given full time traces of the SEIR compartments, we estimate the time derivatives $dF/dt$ for $F = S, E, I, R$ using numpy.gradient~\citep{harris2020array} (which calculates first differences at the boundaries of the domain and central differences in the interior). We then rearrange the formulae Eq~\ref{eq:seir-model} to produce time-varying estimates for $\beta$, $\kappa$ and $\gamma$.

\begin{figure}
    \centering
    \includegraphics[width=0.9\linewidth]{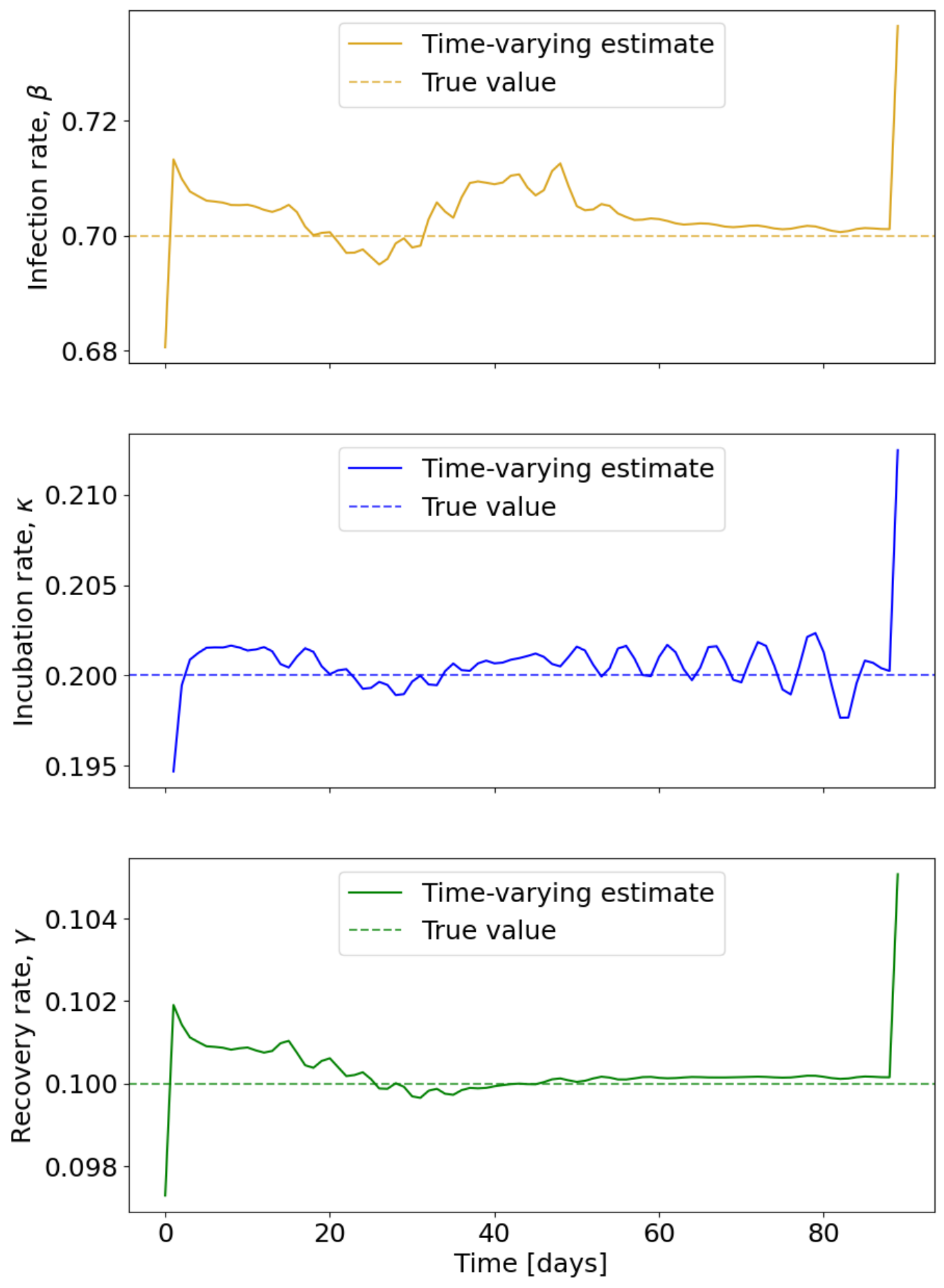}
    \caption{{\bf Time-varying estimates of epidemiological parameters with fixed true values.} We simulated the compartmental ODE model Eq~\ref{eq:seir-model} with $\beta = 0.7$, $\kappa = 0.2$ and $\gamma = 0.1$. We then calculated the time-varying estimates described in Section 6 and plotted them.}
    \label{fig:seirmo-validation}
\end{figure}

To validate this method of estimation, we used the Seirmo package~\citep{seirmo2020} to generate SEIR compartmental data with known epidemiological parameters (Fig~\ref{fig:seirmo-validation}). While these estimates do vary in time, the deviation from their true values is minimal. This is in contrast to Fig 2B and Fig 4B of the main text, in which these three time-varying estimates vary substantially in comparison.

\subsection*{7 - Fitting the ODE model to data for all SEIR compartments for the smaller infection radius}

\begin{figure}
    \centering
    \includegraphics[width=1.0\linewidth]{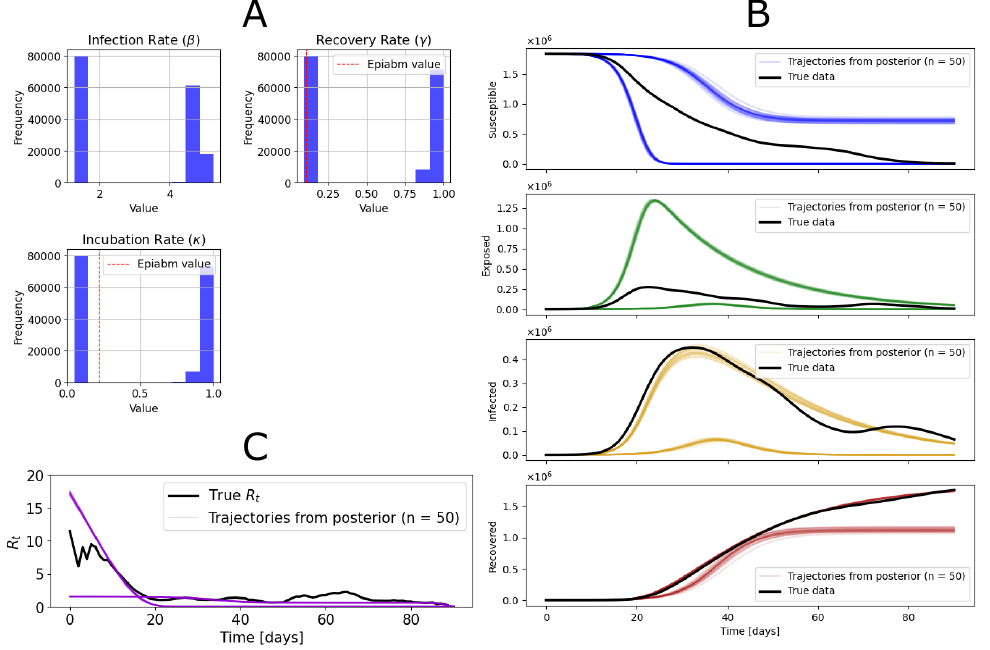}
    \caption{{\bf Inference results for fitting the ODE model to all SEIR compartments with the smaller infection radius.} We fitted the model to the Susceptible, Exposed, Infected and Recovered compartments to infer marginal distributions for $\beta$, $\gamma$ and $\kappa$. A: Samples from the marginal distributions for $\beta$, $\gamma$ and $\kappa$. B: Cloud of inferred time series for the SEIR compartments. We sampled 50 triples ($\beta$, $\kappa$, $\gamma$) from the posterior distribution and simulated the model. From here, we plotted the individual traces for each compartment. C: Cloud of inferred time series for the $R_{t}$ trace. We used the 50 triples ($\beta$, $\kappa$, $\gamma$) from B and calculated the case reproduction number using the method described in Section 2.2.2 of the main text.}
    \label{fig:r0_2-seir-outputs}
\end{figure}

In Section 3.2 of the main text, we presented the results of fitting our ODE model to the prevalence data only. We found that in the case of a smaller infection radius, the compartmental curves generated from forward-simulating the model with the posterior distribution in Fig~\ref{fig:posteriors}B poorly matched the true data from our agent-based model. Following this, we fitted our ODE model to all four model compartments (Susceptible, Exposed, Infected and Recovered) to explore whether this would improve the prediction accuracy. The posterior samples are shown in Fig~\ref{fig:r0_2-seir-outputs}A, the inferred SEIR curves are in Fig~\ref{fig:r0_2-seir-outputs}B and the inferred case $R_{t}$ is shown in Fig~\ref{fig:r0_2-seir-outputs}C. The four chains (initialised at different points in parameter space) did not all converge to the same distribution, and Fig~\ref{fig:r0_2-seir-outputs}B-C indicate that there are at least two modes in this case. Therefore, rather than plotting credible intervals, we sampled 50 parameter sets from the posterior to produce clouds of SEIR traces and resultant clouds of $R_{t}^{\mathrm{case}}$ trajectories. Other than the Recovered compartment, we observed significant deviations in the parameter fits from the SEIR data, and neither mode was able to accurately match the true $R_{t}$. Despite this, each mode captures a small part of the true $R_{t}$ curve relatively well.

\subsection*{8 - Sensitivity of renewal model $R_{t}$ predictions to generation time distribution}

In the main text Section 3.3, we show accurate estimates for $R_{t}$ based on the renewal model inference framework, however these are highly sensitive to the underlying generation time distribution. While we are able to extract realized generation intervals from Epiabm, this is not possible in many real-world inference contexts, and realized intervals may provide at best a biased indication of the intrinsic generation interval, which may impact the accuracy of the inferred $R_{t}$ values.

\begin{figure}
    \centering
    \includegraphics[width=0.85\linewidth]{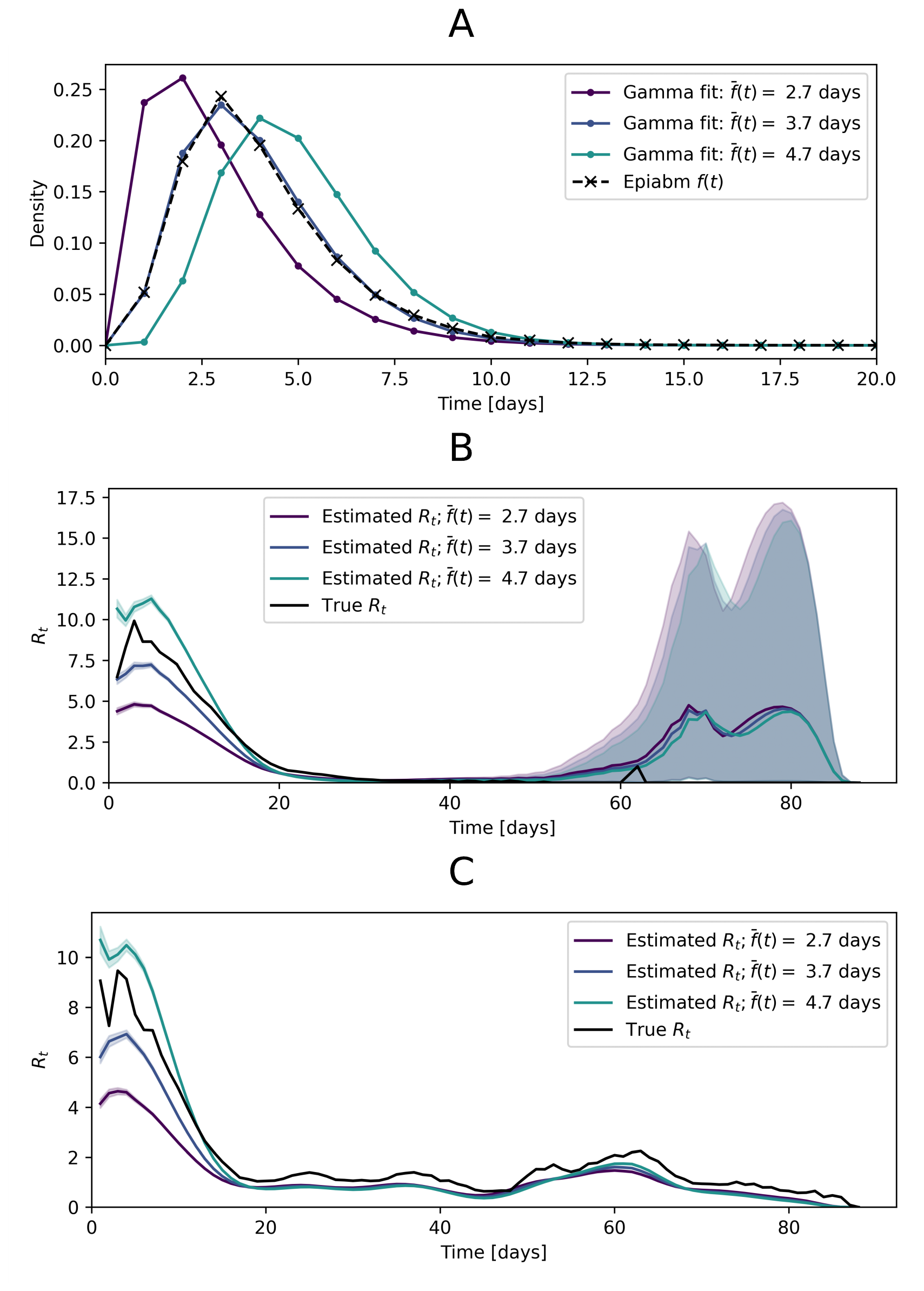}
    \caption{{\bf Sensitivity of \boldmath$R_{t}$ inferred by the renewal model to variation in the generation time \boldmath$f(t)$.} Inferred $R_{t}$ curves are compared as the mean of the generation time is varied. A: A gamma distribution was fitted to the generation time $f(t)$ extracted from Epiabm (with a mean $\bar{f}(t)$ = 3.7 days), in addition to gamma distributions with different means (while the variance is held constant). B, C: The mean posterior $R_{t}$, and 95\% credible interval (shaded) for an infection radius $r = 1$ (B) and an infection radius $r = 0.2$ (C), as each gamma distribution is used as the generation time. }  
    \label{fig:generation_time_sensitivity}
\end{figure}

To exemplify this, we fit the the generation time distribution extracted from Epiabm to a gamma distribution, which we then used as the generation time in the inference framework. In Fig \ref{fig:generation_time_sensitivity} we vary the mean of this gamma distribution, while keeping the variance constant, and show that this significantly affects the accuracy of the $R_{t}$ prediction. 

While changes to the mean of the generation time induce a specific change in the profile of $R_{t}$, other $R_{t}$ dynamics may be captured by further perturbations to the generation time. To exemplify this, we assume the generation time distribution takes the shape of a shifted gamma function, and search for the parameter set that results in the most accurate $R_{t}$ inference result. This accuracy is determined by the lowest root mean squared error between the posterior mean $R_{t}$ and the ground truth $R_{t}$ series, and so this approach would not typically be feasible in wider inference problems, where the ground truth is unknown. 

\begin{figure}
    \centering
    \includegraphics[width=0.80\linewidth]{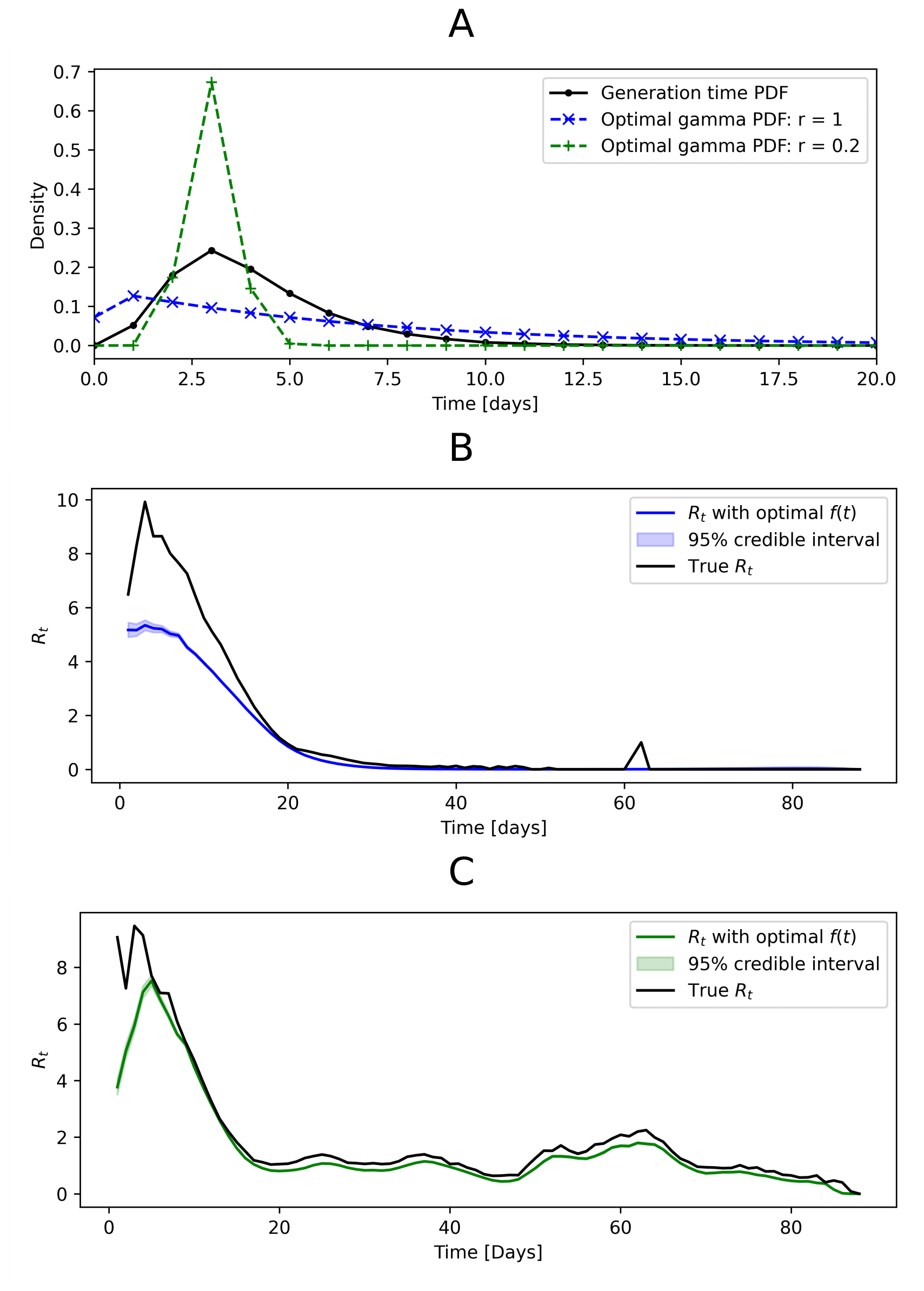}   
    \caption{{\bf Inferred \boldmath$R_{t}$ curves, where the gamma distribution is chosen to obtain the best possible fit to the true \boldmath$R_{t}$.} The parameters for the shifted gamma function were varied, to determine the distribution that results in the lowest root mean squared (rms) error in the $R_{t}$ prediction. A: The optimised generation time distributions, compared to the distribution extracted from the Epiabm data. B: The posterior mean $R_{t}$ curve for the best-fit generation time on the simulation data with an infection radius $r = 1$. Note that this optimised generation time does not predict $R_{t}$ more accurately than the generation time distribution extracted from Epiabm at early times, but does correctly predict a zero $R_{t}$ at later times, reducing the rms error. C: The posterior mean $R_{t}$ curve for the $r = 0.2$ simulation case. }
    \label{fig:generation_time_optimisation}
\end{figure}

Moreover, we do not suggest that this optimal generation time distributions should be interpreted as a `more accurate' representation of the underlying infection dynamics than the generation time than the distribution extracted from Epiabm. The two optimal distributions are not consistent between datasets, with the $r = 1$ case having a much flatter profile, seemingly to prevent the prediction of a second wave resurgence in 
$R_{t}$, whereas the much sharper distribution for $r = 0.2$ can capture rapid changes in $R_{t}$. Rather, the results in Fig \ref{fig:generation_time_optimisation} are intended to exemplify the qualitative ways in which the generation function affect the inferred $R_{t}$ profile.

\paragraph*{Video 1}
\label{video:1}{Animation of the simulation with a larger infection radius in Northern Ireland.}

\paragraph*{Video 2}
\label{video:2}{Animation of the simulation with a smaller infection radius in Northern Ireland.}